\documentclass[twocolumn,aps,prl,superscriptaddress]{revtex4-2}
\usepackage{amssymb} 
\usepackage{amsmath,bm} 
\usepackage{graphicx} 
\usepackage[normalem]{ulem} 
\usepackage{multirow} 
\usepackage[colorlinks,linkcolor=blue,urlcolor=blue,citecolor=blue]{hyperref} 
\usepackage{lipsum} 
\usepackage[usenames, dvipsnames]{xcolor} 
\usepackage{tensor} 
\usepackage{isotope} 
\usepackage{amsmath} 
\usepackage{booktabs} 
\usepackage{xspace}
\usepackage{epsfig} 
\usepackage{fancyhdr}
\usepackage{pslatex}   
\usepackage{xcolor}
\usepackage{color} 

\usepackage{lineno}

\allowdisplaybreaks[4] 

\renewcommand{\sout}{\bgroup \color{red} \ULdepth=-.5ex \ULset}

\begin{document} 

\title{Shedding Light on (Anti-)nuclei Production with Pion-Nucleus Femtoscopy}

\author{Li-Yuan Zhang} 
\affiliation{Key Laboratory of Nuclear Physics and Ion-beam Application~(MOE), Institute of Modern Physics, Fudan University, Shanghai $200433$, China} 
\affiliation{Shanghai Research Center for Theoretical Nuclear Physics, NSFC and Fudan University, Shanghai 200438, China}

\author{Ze-Hua Zhang} 
\affiliation{Key Laboratory of Nuclear Physics and Ion-beam Application~(MOE), Institute of Modern Physics, Fudan University, Shanghai $200433$, China} 
\affiliation{Shanghai Research Center for Theoretical Nuclear Physics, NSFC and Fudan University, Shanghai 200438, China}

\author{Che Ming Ko} 
\email{ko@comp.tamu.edu} 
\affiliation{Cyclotron Institute and Department of Physics and Astronomy, Texas A\&M University, College Station, Texas 77843, USA} 

\author{Yu-Gang Ma} 
\email{mayugang@fudan.edu.cn} 
\affiliation{Key Laboratory of Nuclear Physics and Ion-beam Application~(MOE), Institute of Modern Physics, Fudan University, Shanghai $200433$, China} 
\affiliation{Shanghai Research Center for Theoretical Nuclear Physics, NSFC and Fudan University, Shanghai 200438, China} 

\affiliation{School of Physics, East China Normal University, Shanghai 200241, China}
\author{Qi-Ye Shou} 
\email{shouqiye@fudan.edu.cn} 
\affiliation{Key Laboratory of Nuclear Physics and Ion-beam Application~(MOE), Institute of Modern Physics, Fudan University, Shanghai $200433$, China} 
\affiliation{Shanghai Research Center for Theoretical Nuclear Physics, NSFC and Fudan University, Shanghai 200438, China}

\author{Kai-Jia Sun} 
\email{kjsun@fudan.edu.cn} 
\affiliation{Key Laboratory of Nuclear Physics and Ion-beam Application~(MOE), Institute of Modern Physics, Fudan University, Shanghai $200433$, China} 
\affiliation{Shanghai Research Center for Theoretical Nuclear Physics, NSFC and Fudan University, Shanghai 200438, China}

\author{Zhan-Duo Tang} 
\email{zdtang@fudan.edu.cn}
\affiliation{Key Laboratory of Nuclear Physics and Ion-beam Application~(MOE), Institute of Modern Physics, Fudan University, Shanghai $200433$, China} 
\affiliation{Shanghai Research Center for Theoretical Nuclear Physics, NSFC and Fudan University, Shanghai 200438, China}

\author{Rui Wang}   
\email{rwang@phy.ecnu.edu.cn}  
\affiliation{School of Physics, East China Normal University, Shanghai 200241, China}

\author{Song Zhang} 
\email{song\_zhang@fudan.edu.cn} 
\affiliation{Key Laboratory of Nuclear Physics and Ion-beam Application~(MOE), Institute of Modern Physics, Fudan University, Shanghai $200433$, China} 
\affiliation{Shanghai Research Center for Theoretical Nuclear Physics, NSFC and Fudan University, Shanghai 200438, China}
\date{\today} 

\begin{abstract} 
High-energy nuclear collisions provide a unique environment for synthesizing both nuclei and antinuclei (such as $\bar{d}$ and $^4\overline{\text{He}}$) at temperatures ($k_BT\sim100$ MeV) much higher than their binding energies per nucleon  of a few MeV. The underlying production mechanism, whether through statistical hadronization, nucleon coalescence, or dynamical regeneration and disintegration, remains unsettled. Here we address this question using  pion-nucleus femtoscopy. By solving relativistic kinetic equations for pion-catalyzed reactions ($\pi NN \leftrightarrow \pi d$) for deuteron production and including final-state $p-$wave scatterings derived from an established effective interaction, we successfully reproduce the resonance peaks of both $\pi^+-p$ and $\pi^+-d$ femtoscopic correlations    observed in  $pp$ collisions at $\sqrt{s} = 13~\mathrm{TeV}$. The interplay between $\Delta$ resonance and $p$-wave scatterings shifts both correlation peaks downward by about $70\text{ MeV}$ relative to  vacuum $\Delta$ decay.  Conversely, both the nucleon coalescence model and the statistical hadronization model significantly underestimate the data and produce additional dips  that are absent from the data. These results provide compelling evidence that pion-catalyzed reactions play a dominant role in the production of light (anti-)nuclei in high-energy nuclear collisions and cosmic rays.
\end{abstract} 

\maketitle 

\emph{Introduction.}{\bf ---}
Loosely-bound light nuclei and antinuclei have been measured in high-energy nuclear collisions at Relativistic Heavy Ion Collider (RHIC) and the Large  Hadron Collider (LHC)~\cite{STAR:2010gyg,STAR:2011eej,STAR:2023fbc,ALICE:2024djx}. This phenomenon, known as `little-bang' nucleosynthesis, has attracted renewed interest because of its relevance to matter–antimatter asymmetry tests~\cite{ALICE:2015rey},   quark-gluon plasma (QGP) bulk properties~\cite{JETSCAPE:2022cob}, and indirect dark matter searches~\cite{ vonDoetinchem:2020vbj, ALICE:2022zuz}.  Despite decades of study, the microscopic mechanism by which such fragile (anti-)nuclei form and survive the hot and dense hadronic environment remains unresolved~\cite{Butler:1963pp,Kapusta:1980zz,Csernai:1986qf,Braun-Munzinger:2018hat,Chen:2018tnh,Ono:2019jxm,Braaten:2024cke}. 

\begin{figure*}[!t]
\includegraphics[width=0.85\textwidth, trim={10 0 0 0}]{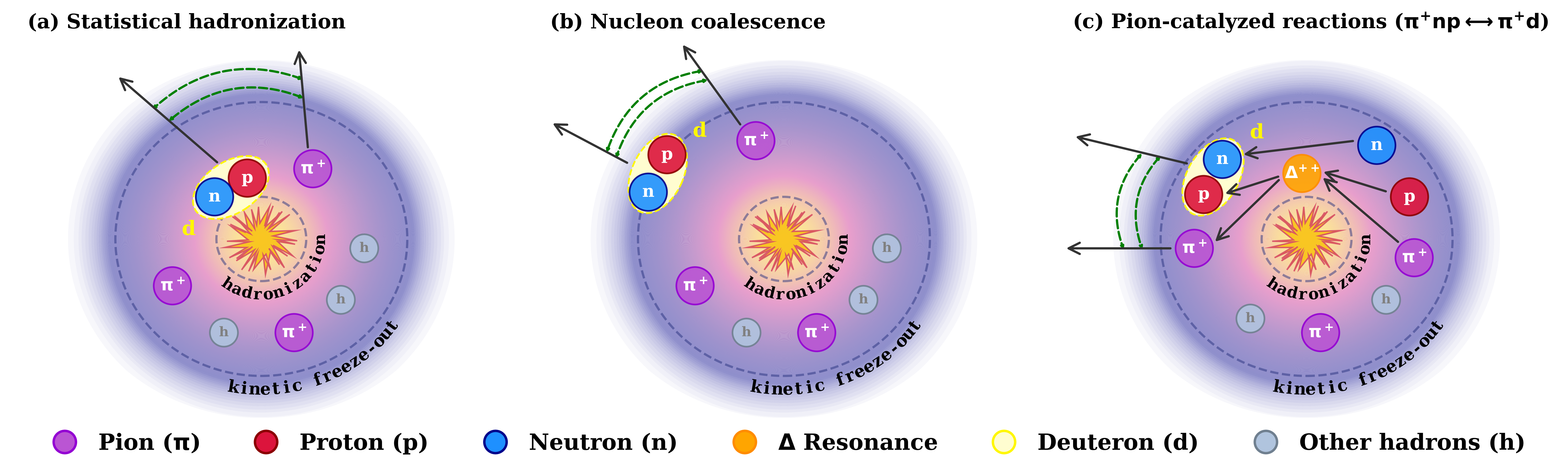}
  \caption{Scenarios of deuteron production in high-energy nuclear collisions, including  statistical hadronization (a),  nucleon coalescence at  kinetic freeze-out  (b), and  pion-catalyzed reactions (e.g. $\pi^+np\leftrightarrow \pi^+d$) with the assistance of intermediate $\Delta$ resonance (c).}
\label{pic:scenario}
\end{figure*}

Three major competing scenarios (see Figure~\ref{pic:scenario})  have been proposed, including statistical hadronization model (SHM)~\cite{Andronic:2017pug}, nucleon coalescence at kinetic freeze-out~\cite{Scheibl:1998tk,Bellini:2020cbj}, and continuous disintegration and regeneration driven by pion-catalyzed reactions in the hadronic phase (e.g. $\pi NN\leftrightarrow\pi d$, $\pi Nd\leftrightarrow\pi^3\mathrm{He}$)~\cite{Oliinychenko:2018ugs,Sun:2022xjr}. Bulk observables such as yields, $p_T$ spectra, and collective flow~\cite{PHENIX:2004vqi,ALICE:2021mfm,ALICE:2022ugx,STAR:2022hbp,STAR:2022fnj} alone cannot unambiguously distinguish these scenarios, as different models can reproduce similar integrated yields and flow patterns~\cite{Oliinychenko:2018ugs,ALICE:2020chv}.

In contrast, correlation femtoscopy~\cite{Lednicky:2002fq,Lisa:2005dd} or Hanbury-Brown and Twiss (HBT) interferometry~\cite{HanburyBrown:1956bqd,Baym:1997ce,Heinz:1999rw} has evolved from its origins in astronomical intensity interferometry  into a precision tool for probing the spatio-temporal structure of relativistic nuclear collisions, with sensitivity to  length scales of order $10^{-15}\,\mathrm{m}$ and time scales of order $10^{-23}\,\mathrm{s}$~\cite{Lednicky:1981su}. It has also become a powerful tool for studying the nature of strong interactions~\cite{STAR:2015kha,ALICE:2020mfd,ALICE:2019gcn,Si:2025eou} and exotic hadrons~\cite{Liu:2024uxn}. Nontrivial femtoscopic correlations can  arise from quantum statistics~\cite{Koonin:1977fh},  final-state strong and Coulomb interactions~\cite{Koonin:1977fh,Wiedemann:1996ig}, or resonance decays~\cite{Wiedemann:1996ig}. Beyond its conventional applications to hadron–hadron pairs~\cite{Fabbietti:2020bfg},   femtoscopy has recently been extended to hadron–nucleus systems such as proton-deuteron ($p-d$)~\cite{ALICE:2023bny,Mrowczynski:2025qys}, kaon-deuteron ($K-d$)~\cite{VazquezDoce:2024nye}, and deuteron-deuteron ($d-d$)~\cite{Mrowczynski:2021bzy} pairs. For pion-catalyzed channels such as $\pi^+np\to\pi^+d$, the reaction rate peaks through formation of an intermediate $\Delta(1232)$ resonance~\cite{Sun:2022xjr}. This would produce a characteristic resonant enhancement in the pion-deuteron ($\pi-d$) correlation function, an effect not present in $p-d$ or $K-d$ correlations~\cite{ALICE:2023bny}. Such a resonance structure has recently been observed in both $\pi^\pm-p$ and $\pi^\pm-d$ correlations by the ALICE Collaboration~\cite{ALICE:2025aur,ALICE:2025byl}, though its microscopic origin remains to be clarified.

In the present study,  we elucidate the ALICE measurements~\cite{ALICE:2025aur,ALICE:2025byl} by solving relativistic kinetic equations for pion-catalyzed regeneration and breakup of (anti-)deuterons in $pp$ collisions at $\sqrt{s}=13$~TeV and calculating  the $\pi^+-p$ and $\pi^+-d$ correlation functions. These correlation functions are found to exhibit pronounced  peaks at  $\Delta$ resonance mass, corresponding to $k^*\simeq 0.22$ GeV/$c$. Further including final-state $p-$wave scatterings,  both peaks shift downward to $k^*\simeq 0.15$ GeV/$c$, which successfully reproduce the ALICE data.  In contrast, both the nucleon coalescence model and the statistical hadronization model significantly underestimate the data and generate unobserved dips in the correlation functions.  These results demonstrate the unique discriminating power of pion–nucleus femtoscopy and identify pion-catalyzed reactions as the dominant mechanism for the production of light (anti-)nuclei in high-energy nuclear collisions.

\emph{Theoretical framework.}{\bf ---} We employ a multi-phase transport (AMPT) model~\cite{Lin:2004en} to describe the evolution of the partonic phase created in $pp$ collisions, with hadronization implemented via a quark coalescence mechanism.   For the subsequent hadronic evolution, we include not only pion-catalyzed deuteron formation and dissociation but also other relevant scattering channels available in AMPT.   The AMPT model has been widely adopted for studying  medium response to energetic jets~\cite{Ma:2010dv,Luo:2021voy}, particle production~\cite{Oh:2009gx,Shao:2022eyd}, nuclear structure effects \cite{Wang:2021ghq}, and etc.,  in relativistic heavy-ion collisions.

Relativistic kinetic equations have long been  used to describe light  (hyper-)nuclei production in heavy-ion collisions~\cite{Danielewicz:1991dh,Oliinychenko:2018ugs,Sun:2022xjr,Wang:2025wim}.  The  contribution of $\pi^+d\leftrightarrow \pi^+np$  to deuteron production and dissociation in nuclear collisions can be included in the relativistic kinetic equation as~\cite{Sun:2022xjr}
\begin{eqnarray} 
\frac{\partial f_d}{\partial t}+\frac{ {\bf P}}{\ {E_d}}\cdot \frac{\partial f_d}{\partial {\bf R}}=-\mathcal{K}^{>}f_d + \mathcal{K}^{<}(1+f_d), 
\label{eq:deu_kin} 
\end{eqnarray}
where  $f_d({\bf R},{\bf P})$ denotes the phase-space distribution of deuterons, and $\mathcal{K}^{>}$ and $\mathcal{K}^{<}$ denote, respectively, the dissociation and regeneration rates with their expressions given in Ref.~\cite{Sun:2022xjr}.  We neglect in the above equation the mean-field effect on the propagation of deuteron due to its much larger kinetic  than potential energy in the produced hadronic matter. 

Equation (\ref{eq:deu_kin}) has   successfully describe{d (anti-)deuteron production in Au+Au and Pb+Pb collisions~\cite{Oliinychenko:2018ugs,Sun:2022xjr}. It was solved by using  test particles for particle phase-space distribution functions and the stochastic method to treat the collision integrals on the right hand side of Eq.(\ref{eq:deu_kin})~\cite{Sun:2022xjr}. In this method, the probability that the reaction $\pi^+d \rightarrow \pi^+np$, which is treated using the impulse (IA) approximation in our study, occurs between a pion and a deuteron within the volume $\Delta V$ during the time interval $\Delta t$  is given by~\cite{Sun:2022xjr,Danielewicz:1991dh,Xu:2004mz,Wang:2020ixf} 
\begin{eqnarray} 
P_{23}\big{|}_\text{IA} \approx F_d{ v}_{\pi^+ p}\sigma_{\pi^+ p\rightarrow \pi^+ p}\frac{\Delta t}{N_\text{test}{\Delta} V} + (p\leftrightarrow n).
\label{eq:deu_p23}
\end{eqnarray} 
In the above equation, the constant $F_d \approx 0.72$ is introduced to describe the measured $\pi^+-d$ dissociation cross section~\cite{ParticleDataGroup:2020ssz} in terms of the $\pi^+-p$ scattering cross section $\sigma_{\pi^+-p}$ in the energy range relevant to our study, ${{v_{\pi^+ p}}}$ denotes the relative speed between the pion and the proton  in the deuteron, and $N_\text{test}$ refers to the number of test particles used for a physical particle in the test-particle method~\cite{Wong:1982zzb}.  

For the cross section $\sigma_{\pi^+-p}$ in Eq.~(\ref{eq:deu_p23}), it is dominated by the $p$-wave interaction through the $\Delta$ resonance and can be calculated using the $\pi N\Delta$ interaction Lagrangian density~\cite{vanHees:2004vt},
\begin{equation}
\mathcal{L}
=
-\frac{f_\pi}{m_\pi}\,
\psi_N
\left[
(\vec S^{\dagger}\!\cdot\!\vec\nabla)
(\vec T^{\dagger}\!\cdot\!\vec\phi_\pi)
\right]
\Psi_\Delta
+ {\rm h.c.}~.
\label{eq:Lag}
\end{equation}
In the above, $\vec\phi_\pi$, $
\psi_N$, and $\Psi_\Delta$ represent the pion, nucleon, and $\Delta$-resonance fields, respectively;  $f_\pi$ is the  $\pi N\Delta$ coupling constant;   $m_\pi$ is the pion mass; and $\vec S$ and $\vec T$ are the spin 1/2 to spin 3/2 and isospin 1/2 to isospin 3/2 transition operators, respectively. The resulting $\pi^+ - p$ cross section is  given by~\cite{vanHees:2004vt}
\begin{equation} 
\sigma(\sqrt{s}) = \frac{E_N^2}{18\pi s} \left( \frac{f_\pi}{m_\pi} \right)^4 \frac{k^4 F^4}{\bigl(\sqrt{s} - m_\Delta\bigr)^2 + \Gamma^2/4},\label{eq:cs}
\end{equation}
where $k$ is the center-of-mass (c.m.) momentum; $E_N=\sqrt{m_N^2+k^2}$ with $m_N$ being the proton mass is the proton energy, $\sqrt{s}=E_N+E_\pi$ with $E_\pi=\sqrt{m_\pi^2+k^2}$ is the total c.m. energy; $m_\Delta$ and $\Gamma$ are the mass and width of the $\Delta$ resonance, respectively; and  $F(k)=\Lambda^2/(\Lambda^2+k^2)$ is the form factor at the $\pi N\Delta$ vertex. With $m_{\pi}=0.139$ GeV, $m_{N}=0.939$ GeV, $m_{\Delta}=1.232$ GeV, $f_\pi=3.3$,  $\Lambda=0.33$ GeV, and $\Gamma$ evaluated using Eq.~(\ref{eq:Lag}),  Eq.~(\ref{eq:cs}) can accurately reproduce the experimentally measured $\pi^+-p$ cross section~\cite{ParticleDataGroup:2020ssz}. 

\begin{figure*}[!t]
  \centering 
 \includegraphics[width=14.6cm]{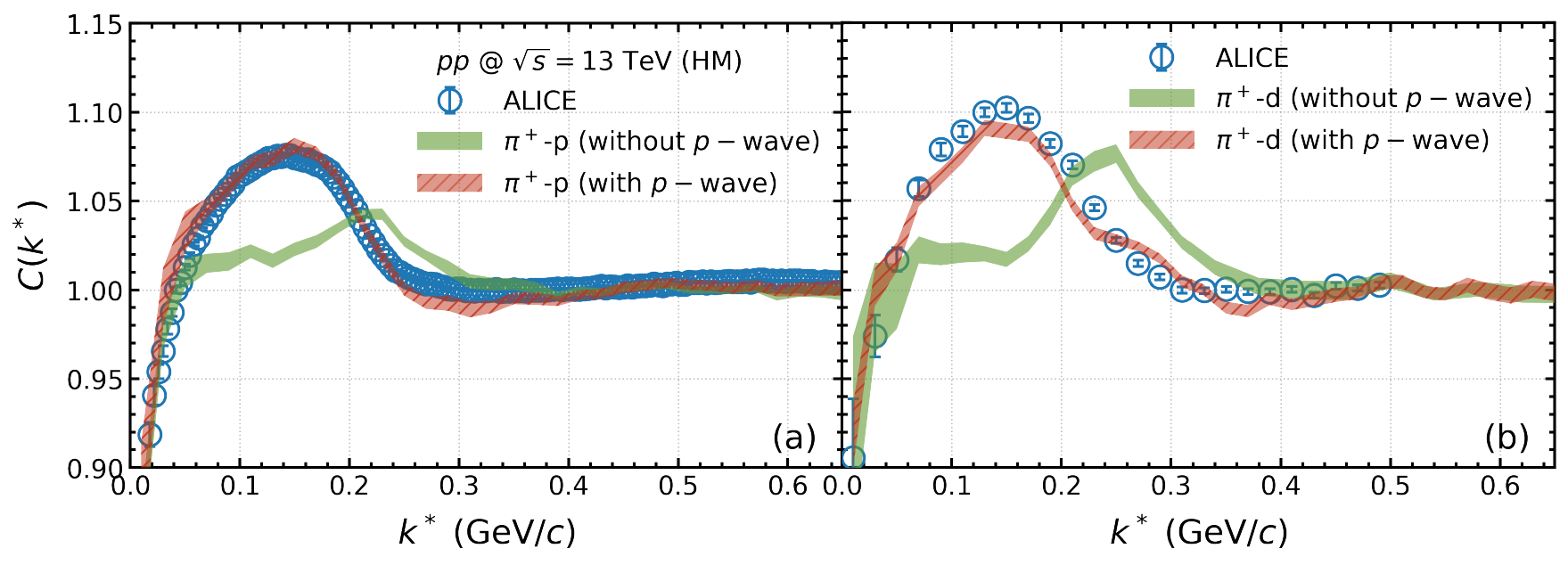}
  \caption{Correlation function $C(k^*)$ of $\pi^+-p$ (a) and $\pi^+-d$ (b) in high-multiplicity $pp$ collisions at $\sqrt{s}=13$ TeV. Theoretical results with  and without $p-$wave scatterings are shown by red and green  bands, respectively.  Experimental data taken from the ALICE Collaboration~\cite{ALICE:2025aur,ALICE:2025byl} are denoted by filled symbols, where particle rapidity is restricted to $|\eta| \leq 0.8$.  For $\pi^+-p$ correlations, an additional cut of pair transverse momentum, defined by  $m_T=\sqrt{(m_p+m_\pi)^2+(\mathbf{k}_p+\mathbf{k}_\pi)^2}/2$, of $m_T\in$ [0.54 0.75) GeV/$c^2$ is imposed~\cite{ALICE:2025aur}.} 
  \label{pic:MB}
\end{figure*}


For the convergence of numerical results from the solution of the kinetic equation, a sufficiently large $N_\text{test}$ is  required. However, calculating femtoscopic correlation functions necessitates event-by-event simulations ($N_\text{test} = 1$), as any  $N_\text{test} > 1$ artificially suppresses genuine particle correlations. To overcome this limitation, we replace the $1/\Delta V$ factor in Eq.~(\ref{eq:deu_p23}) with a Gaussian smearing function, 
$S_{ij} =  e^{-d_{ij}^2 / (2l^2)}/(2\pi l^2)^{\frac{3}{2}}$, where $d_{ij}$ is the spatial distance between particles $i$ and $j$, and $l$ denotes the effective interaction range. This treatment mimics the scattering of two Gaussian wave packets~\cite{Ono:2019jxm}. With this modification, Eq.~(\ref{eq:deu_p23}) is rewritten as 
\begin{eqnarray} 
P_{23}\big{|}_\text{IA} \approx F_d{ v}_{\pi^+ p}\sigma_{\pi^+ p\rightarrow \pi^+ p}S_{\pi^+p}\Delta t + (p\leftrightarrow n).
\label{eq:deu_p23_new}
\end{eqnarray} 
Since the value of $l$ does not affect the momentum correlations of  outgoing particles, we  set  $l = 1.2~\mathrm{fm}$ in the following calculation. Similarly, the probability for the reaction $\pi^+np \rightarrow \pi^+d$ to take place   in a time interval $\Delta t$ is 
\begin{eqnarray} 
P_{32}\big{|}_\text{IA} \approx g_dF_d{ v}_{\pi^+p}\sigma_{\pi^+ p\rightarrow \pi^+ p} S_{\pi^+p} W_d\Delta t  +(p\leftrightarrow n),
\label{eq:deu_p32_wig}
\end{eqnarray} 
where the $g_d=$3/4 is the statistical factor for spin 1/2 proton and neutron to form a spin 1 deuteron, and $W_d$ stands for the deuteron Wigner function, which depends on the relative position and momentum of the proton and neutron in the reaction.

The kinetic approach based on Eqs.~(\ref{eq:deu_p23_new}) and (\ref{eq:deu_p32_wig}) assumes instantaneous deuteron formation upon pion–nucleon interaction. This assumption is supported by recent quantum-mechanical calculations~\cite{Rais:2022gfg}, which show  negligible  formation time delay of  deuteron production, contrary to the claim in Ref.~\cite{Mrowczynski:2019yrr}.  It is further assumed in the kinetic approach that the deuteron remains bound in the produced hadronic matter, as the high temperature and low nucleon density in high-energy nuclear collisions suppress its Mott dissociation due to Pauli blocking~\cite{Typel:2009sy}.  In contrast, the Mott effect becomes important  in  the low temperature and high nucleon density environment in intermediate-energy nuclear collisions~\cite{Wang:2023gta,Wang:2025wim}.

Using particle phase-space distributions at kinetic freeze-out, the two-particle correlation function is calculated as~\cite{Lisa:2005dd}
\begin{eqnarray}
C(\mathbf{k^*}) = \mathcal{N}\frac{\sum_{\text{pairs}}^{\text{same}} \delta(\mathbf{k^*_{\text{pair}}} - \mathbf{k^*})   w(\mathbf{k^*},\mathbf{r^*})}{\sum_{\text{pairs}}^{\text{mixed}} \delta(\mathbf{k^*_{\text{pair}}} - \mathbf{k^*})}, \label{eq:CF2}
\end{eqnarray}
where $\mathbf{k}^* = (\mathbf{k}_2 - \mathbf{k}_1)/2$ and $\mathbf{r}^* = \mathbf{r}_2 - \mathbf{r}_1$ are the relative momentum and spatial separation in the pair rest frame  with  the earlier-emitted particle  freely propagated to the emission time of the later produced one~\cite{Kisiel:2009eh}.  The numerator and denominator represent the same-event and mixed-event pair relative-momentum distributions, respectively, while $\mathcal{N}$ normalizes $C(\mathbf{k}^*) \to 1$ at large $k^*$ where the two-body correlation vanishes. The same-event weight function $w(\mathbf{k}^*,\mathbf{r}^*)$ in Eq. (\ref{eq:CF2}) is given by $w(\mathbf{k}^*,\mathbf{r}^*) = \vert{}\Psi_{\mathbf{k}^*}(\mathbf{r}^*)\vert{}^2$  with $\Psi_{\mathbf{k}^*}(\mathbf{r}^*)$ denoting the scattering wave function of the two particles due to their final-state interactions (FSI), i.e.,  $\Psi_{\textbf{k}^{*}}(\textbf{r}^{*}) = e^{i\textbf{k}^{*}\cdot \textbf{r}^{*}} + (2\pi)^{-3}\int  d^{3}{\textbf{k}'}  G(\textbf{k}^{*},\textbf{k}')T({\bf k}^\prime, \textbf{k}^{*})e^{i\boldsymbol{k'}\cdot \textbf{r}^{*}}$, where  $G$ and $T$  are the two-particle propagator and $T-$matrix, respectively.  

For the $\pi^+-p$ correlation,  the weight function $w({\bf k}^*, {\bf r}^*)$ is obtained from $\left|\Psi_{{\bf k}^*}({\bf r}^*)\right|^2$  after averaging over initial  and summing over final spin states, and it is given by~\cite{Vidana:2023olz,Liu:2024nac}
\begin{eqnarray} 
&&\overline{|\Psi_{{\bf k}^*}({\bf r}^*)|^2} = 1 + |U(k^*,r^*)|^2 \left( \frac{1}{3} \cos^2 \theta + \frac{1}{9} \right) \notag\\
&&\hspace{2cm}+ \frac{4}{3} \cos \theta \, \text{Re} \left( U(k^*, r^*) e^{-i\textbf{k}^{*} \cdot \textbf{r}^{*}} \right), \label{eq:wf}
\end{eqnarray}
with 
\begin{eqnarray}\label{eq:amplitude} 
&&U(k^*, r^*) = \frac{ik}{2\pi^2}\frac{f_\pi^2}{m_\pi^2} \frac{F(k^*)m_\pi m_N}{\sqrt{2E_\pi(k^*)} \left( \sqrt{s} - m_\Delta + i\Gamma(k^*)/2 \right)} \notag \\
&&\hspace{0.5cm}\times \int_0^\infty  \frac{  (2E_\pi(k'))^{-\frac{1}{2}}}{E_\pi(k') E_N(k')}\frac{ k'^3 F(k') j_1(k'r^*) dk'}{(\sqrt{s} - E_N(k') - E_\pi(k') + i\epsilon)} ,
\end{eqnarray}
where $j_1(k^\prime r)$ is the spherical Bessel function. The last term in Eq.~(\ref{eq:wf}) denotes the interference between the   initial plane wave and the scattered wave. This interference term has a peak below the $\Delta$ mass and a dip on the right side of the peak. 

For $\pi^+-d$ scattering treated in the impulse approximation, the function $U(k^*,r^*)$ in Eq.~(\ref{eq:wf}) is replaced by $\sqrt{4/3}U(\tilde{k}^*,\tilde{r}^*)$, where $\tilde{k}^*$ and $\tilde{r}^*$ are the relative momentum and position between the  pion and the proton inside the deuteron.  The  factor $\sqrt{4/3}$ accounts for the  cross-section relation $\sigma_{\pi^+-d}\approx\frac{4}{3}\sigma_{\pi^+-p}$. The ${\tilde k}^*$  is determined from the invariant mass of the pion and the proton inside the deuteron, which we take as half of the deuteron momentum.  Neglecting smearing effects due to the deuteron wave function, the $\tilde{r}^*$ is taken  simply  as the pion-deuteron relative distance. For the Coulomb corrections to the correlation function, they are given in the Supplemental Material.  

For $\pi^+$ and $p$ pairs produced directly from the  $\Delta$ resonance decay or $\pi^+$ and $d$ pairs originating from pion-catalyzed reaction, they do not undergo FSI.  Their same-event weight function is taken to be a constant, i.e., $w(\mathbf{k}^*,\mathbf{r}^*) = \alpha_r$, with   its value fitted to the measured $\pi^+-p$  correlation function.  

\emph{Pion-nucleus femtoscopy  in $pp$ collisions at $\sqrt{s}=13$ TeV.}{\bf ---} We  consider the pion–nucleus femtoscopic correlation functions in high‑multiplicity (HM)  $pp$ collisions at $\sqrt{s}=13\ \mathrm{TeV}$~\cite{ALICE:2019buq}.  Figure~\ref{pic:MB} displays the correlation functions $C(k^*)$ for   $\pi^+-p$ (a) and $\pi^+-d$ (b)  pairs from theoretical calculations as well as their comparison with the ALICE data ~\cite{ALICE:2025aur,ALICE:2025byl} denoted by filled symbols together with their combined statistical and systematic uncertainties.  

Without final-state $p$-wave scattering, the theoretical  results for events with charged particle multiplicity $\langle dN_\text{ch}/d\eta\rangle \approx 30$, shown by green bands, display a pronounced peak at $k^*\approx0.22~\mathrm{GeV}/c$.  This peak  in the $\pi^+-p$ correlation function originates from the $\Delta^{++}\rightarrow \pi^++p$ decay. For the $\pi^+-d$  correlation function, it stems from the forward pion-catalyzed reaction ($\pi NN\rightarrow\pi d$) through the formation of an intermediate $\Delta$ resonance~\cite{Sun:2022xjr} or, equivalently, via the sequential $\Delta$-assisted processes $\pi N\leftrightarrow\Delta$ and $\Delta N\leftrightarrow\pi d$. In both $\pi^+-p$ and $\pi^+-d$ correlation functions, $C(k^*)$ is suppressed at low relative momentum ($k^*\le0.05~\mathrm{GeV}/c$) by   Coulomb repulsion.

Also shown in Fig.~\ref{pic:MB} are $\pi^+-p(d)$   correlation functions measured by the ALICE Collaboration, which show, however, peaks  at a smaller  $k^* \approx 0.15~\mathrm{GeV}/c$. This downward shift of the peak  was initially interpreted as  the reflection of an unexpectedly low freeze-out temperature ($k_BT \approx 20\text{--}30\text{ MeV}$)~\cite{ALICE:2025aur,ALICE:2025byl} well below standard values ($>100\text{ MeV}$) in high-energy $pp$ collisions~\cite{ALICE:2020nkc}. Analogous $\Delta$ mass shifts have been reported in $pp$ and $d$+Au collisions at $\sqrt{s_{NN}}=200~\mathrm{GeV}$~\cite{STAR:2008twt}, Au+Au collisions at $\sqrt{s_{NN}}=2.4~\mathrm{GeV}$~\cite{Adamczewski-Musch:2020edy}, and earlier fixed-target experiments~\cite{FOPI:1998clk,Hjort:1997wk}. At lower energies ($\sqrt{s_{NN}} < 3\text{ GeV}$), such modifications were attributed to $\Delta$ regeneration near freeze-out~\cite{Reichert:2019lny,Reichert:2020uxs} and its thermalization  by cold spectator nuclear matter~\cite{FOPI:1998clk,Weinhold:1996ts}.

Incorporating $p$-wave scattering  drastically improves  the  agreement of theoretical results with the ALICE data. As shown by the red band in Fig.~\ref{pic:MB} (a), obtained with the same-event weight function for $\pi^+-p$ pairs from the decay of $\Delta$ resonances taken to be $\alpha_r=1.8$, the peak in the $\pi^+-p$ correlation function shifts to a lower relative momentum $k^*\approx 0.15~\mathrm{GeV}/c$. This effect is due to the interference term in Eq.~(\ref{eq:wf}), which produces a peak below the $\Delta$ mass and is also accompanied by a high-$k^*$ dip, which  is, however, compensated by the contribution from direct $\Delta$-decay  to the $\pi^+-p$ correlation function.  A similar mechanism operates in $\pi^+-d$ correlations, where their $p$-wave scattering likewise yields an excellent  description of the ALICE measurements. 

The fitted value $\alpha_r > 1$ indicates that the contribution from direct $\Delta$-decay   is underestimated in  the AMPT model, warranting further study. The role of $p$-wave scattering in $\pi$-$p$ correlations was also recently examined in Refs.~\cite{Zhang:2026psh,Liu:2026vkc}, with Ref.~\cite{Liu:2026vkc} employing four parameters rather than a single $\alpha_r$.

Finally, in the hybrid FIST+SMASH framework, in which  the statistical model (FIST) is used to produce deuterons at hadronization,  the subsequent hadronic transport in SMASH is dominated by deuteron dissociation, creating a depletion   in the $\pi$-$d$ correlation~\cite{ALICE:2025byl}. In contrast, our approach generates deuterons at post-hadronization stage, where regeneration  dominates and naturally reproduces the observed peaks. 

\begin{figure}[!t]
  \centering 
 \includegraphics[width=8.4cm]{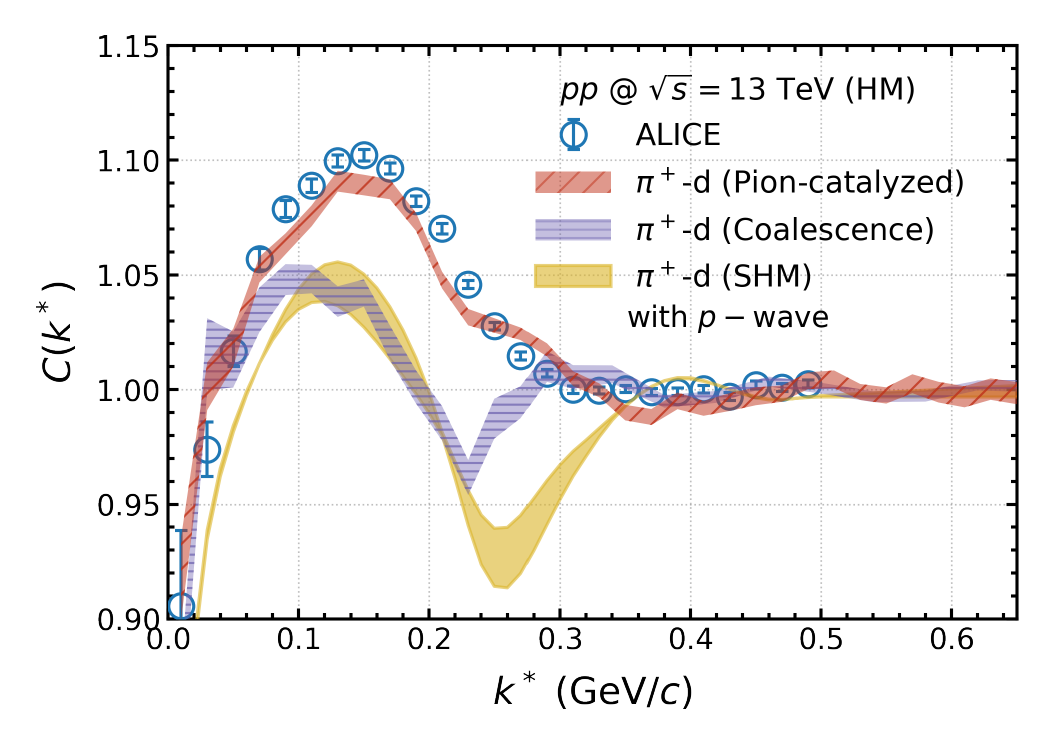} 
  \caption{Correlation function $C(k^*)$ of $\pi^+-d$  in different production scenarios for $pp$ collisions at $\sqrt{s}=13$ TeV.   Experimental data   shown as filled symbols  are taken from the ALICE Collaboration~\cite{ALICE:2025byl}, and theoretical predictions  are represented by colored bands.}
  \label{pic:model_comp}
\end{figure}

\emph{Pion-nucleus femtoscopy in different production scenarios.}{\bf ---} We next investigate the pion–nucleus correlation functions  in the frameworks of the nucleon coalescence model~\cite{Scheibl:1998tk} and the statistical hadronization model~\cite{Andronic:2017pug}.   In the coalescence approach, deuterons form via $n+p\rightarrow d$  at the kinetic freeze-out, and its number ($N_d$) is given by~\cite{Scheibl:1998tk} 
\begin{align}
N_d =g_d \sum_{np}W_{d} &= 6\sum_{np}\exp\!\Bigl(-\tfrac{r^{*2}}{\sigma_d^2} - \sigma_d^2 k^{*2}\Bigr),
\label{Eq:wig}
\end{align}
where the summation is over all  neutron and proton pairs in each collision event. We have taken in the above the deuteron Wigner function $W_d$ to be Gaussian for simplicity, with its width parameter  $\sigma_d = \sqrt{8/3}\,r_d\approx3.2\ \mathrm{fm}$ to reproduce the deuteron  radius $r_d=1.96$ fm~\cite{Ropke:2008qk,Sun:2018mqq}.

The $\pi^+$-$d$ correlation function predicted by the coalescence model is shown by the purple band in Fig.~\ref{pic:model_comp}. Due to $p$-wave scattering, the correlation peak shifts to $k^* \approx 0.12~\mathrm{GeV}/c$, and a small dip  appears at $k^* \approx 0.22~\mathrm{GeV}/c$.   Contrary to the results from pion-catalyzed reactions, the coalescence-induced positive correlation between $\pi^+$ and $d$, when the constituent nucleon originates from the $\Delta$ decay, is insufficient to compensate for the dip produced by the $p$-wave interference, leaving a   depletion on the high-$k^*$ side of the peak. 

In the statistical hadronization model, pions and nuclei are emitted independently from a thermal source of an effective Gaussian  radius of 1.51$\pm 0.12$ fm~\cite{ALICE:2025byl}. With the inclusion of  $\pi^+-d$ $p$-wave  scattering, the resulting correlation function, shown by the yellow band in Fig.~\ref{pic:model_comp},  exhibits a suppressed peak and a pronounced dip owing to the absence of resonance contributions—in contrast to Ref.~\cite{ALICE:2025byl}, where the $p$-wave scattering between $\pi^+$ and $d$ was not included. 
 
The stark differences among results from  the above two theoretical frameworks and those from the kinetic approach demonstrate the exceptional discriminating power of pion–nucleus femtoscopy in identifying the mechanism for light (anti-)nucleus production.

\emph{Conclusion.}{\bf ---}
Using pion–nucleus femtoscopy, we have elucidated the dynamics of light-nucleus production in high-energy $pp$ collisions. Solving the relativistic kinetic equations for pion-catalyzed reactions $\pi NN \leftrightarrow \pi d$ for deuteron production and dissociation during the evolution of produced hadronic matter and including the  final-state $p$-wave scattering between $\pi^+$ and $p$ as well as between $\pi^+$ and $d$, our model reproduces the $\Delta$-resonance peaks in $\pi^+$--$p$ and $\pi^+$--$d$ correlations observed by ALICE at $\sqrt{s} = 13\ \mathrm{TeV}$.  On the other hand, deuteron production based on the coalescence  model or the statistical hadronization model fail to describe the data owing to missing or incomplete  correlations due to the $\Delta$ resonance. These results thus strongly suggest that the  pion-catalyzed reactions  are the dominant mechanism for light (anti-)nuclei production in nuclear collisions at LHC energies.

Whereas the  recent  novel macroscopic approach in terms of  the thermodynamic variable  contact, which is   conjugate to deuteron's binding momentum, can determine the  deuteron  production rate at kinetic freeze-out of high-energy nuclear collisions~\cite{Braaten:2024cke}, the $\pi-d$ femtoscopy   can isolate the specific reaction dynamics for  its production. Extending this method to heavier  nuclei, including $\pi$--${}^3\mathrm{He}$, $\pi$--$\alpha$, and $\pi$--${}^3_\Lambda\mathrm{H}$ correlations, will provide crucial insights into the production mechanisms for light (hyper-)nuclei  in nuclear collisions and cosmic rays, as well as that of exotic hadrons like $X(3872)$ and $T_{cc}^+$ in these collisions.

\emph{Acknowledgments.}{\bf ---} 
We thank Lie-Wen Chen, Mei-Yi Chen, Bhawani Singh,  Laura Fabbietti, Yi-Heng Feng, Li-Sheng Geng, Feng-Kun Guo, Dimitar Mihaylov, Dai-Neng Liu, Pok Man Lo, Xiao-Feng Luo, Dong-Fang Wang, and Zhen Zhang for fruitful discussions.  This work was supported in part by the National Key Research and Development Project of China under Grant No. 2024YFA1612500;  the National Natural Science Foundation of China under Contracts No. 12422509, No. 12375121, No. 12025501, No. 12147101, No. 11891070, No. 124B2102, No. 12275054, and No. 12061141008; Shanghai Pilot Program for Basic Research - Fudan University 21TQ1400100(22TQ006);  the Guangdong Major Project of Basic and Applied Basic Research No. 2020B0301030008; and the STCSM under Grant No. 23590780100. The computations in this research were performed using the CFFF platform of Fudan University.  


\begin{thebibliography}{79}%
\makeatletter
\providecommand \@ifxundefined [1]{%
 \@ifx{#1\undefined}
}%
\providecommand \@ifnum [1]{%
 \ifnum #1\expandafter \@firstoftwo
 \else \expandafter \@secondoftwo
 \fi
}%
\providecommand \@ifx [1]{%
 \ifx #1\expandafter \@firstoftwo
 \else \expandafter \@secondoftwo
 \fi
}%
\providecommand \natexlab [1]{#1}%
\providecommand \enquote  [1]{``#1''}%
\providecommand \bibnamefont  [1]{#1}%
\providecommand \bibfnamefont [1]{#1}%
\providecommand \citenamefont [1]{#1}%
\providecommand \href@noop [0]{\@secondoftwo}%
\providecommand \href [0]{\begingroup \@sanitize@url \@href}%
\providecommand \@href[1]{\@@startlink{#1}\@@href}%
\providecommand \@@href[1]{\endgroup#1\@@endlink}%
\providecommand \@sanitize@url [0]{\catcode `\\12\catcode `\$12\catcode
  `\&12\catcode `\#12\catcode `\^12\catcode `\_12\catcode `\%12\relax}%
\providecommand \@@startlink[1]{}%
\providecommand \@@endlink[0]{}%
\providecommand \url  [0]{\begingroup\@sanitize@url \@url }%
\providecommand \@url [1]{\endgroup\@href {#1}{\urlprefix }}%
\providecommand \urlprefix  [0]{URL }%
\providecommand \Eprint [0]{\href }%
\providecommand \doibase [0]{https://doi.org/}%
\providecommand \selectlanguage [0]{\@gobble}%
\providecommand \bibinfo  [0]{\@secondoftwo}%
\providecommand \bibfield  [0]{\@secondoftwo}%
\providecommand \translation [1]{[#1]}%
\providecommand \BibitemOpen [0]{}%
\providecommand \bibitemStop [0]{}%
\providecommand \bibitemNoStop [0]{.\EOS\space}%
\providecommand \EOS [0]{\spacefactor3000\relax}%
\providecommand \BibitemShut  [1]{\csname bibitem#1\endcsname}%
\let\auto@bib@innerbib\@empty
\bibitem [{\citenamefont {Abelev}\ \emph {et~al.}(2010)\citenamefont {Abelev}
  \emph {et~al.}}]{STAR:2010gyg}%
  \BibitemOpen
  \bibfield  {author} {\bibinfo {author} {\bibfnamefont {B.~I.}\ \bibnamefont
  {Abelev}} \emph {et~al.} (\bibinfo {collaboration} {STAR}),\ }\bibfield
  {title} {\bibinfo {title} {{Observation of an Antimatter Hypernucleus}},\
  }\href {https://doi.org/10.1126/science.1183980} {\bibfield  {journal}
  {\bibinfo  {journal} {Science}\ }\textbf {\bibinfo {volume} {328}},\ \bibinfo
  {pages} {58} (\bibinfo {year} {2010})}\BibitemShut {NoStop}%
\bibitem [{\citenamefont {Agakishiev}\ \emph {et~al.}(2011)\citenamefont
  {Agakishiev} \emph {et~al.}}]{STAR:2011eej}%
  \BibitemOpen
  \bibfield  {author} {\bibinfo {author} {\bibfnamefont {H.}~\bibnamefont
  {Agakishiev}} \emph {et~al.} (\bibinfo {collaboration} {STAR}),\ }\bibfield
  {title} {\bibinfo {title} {{Observation of the antimatter helium-4
  nucleus}},\ }\href {https://doi.org/10.1038/nature10079} {\bibfield
  {journal} {\bibinfo  {journal} {Nature}\ }\textbf {\bibinfo {volume} {473}},\
  \bibinfo {pages} {353} (\bibinfo {year} {2011})},\ \bibinfo {note} {[Erratum:
  Nature 475, 412 (2011)]}\BibitemShut {NoStop}%
\bibitem [{\citenamefont {Abdulhamid}\ \emph {et~al.}(2024)\citenamefont
  {Abdulhamid} \emph {et~al.}}]{STAR:2023fbc}%
  \BibitemOpen
  \bibfield  {author} {\bibinfo {author} {\bibfnamefont {M.}~\bibnamefont
  {Abdulhamid}} \emph {et~al.} (\bibinfo {collaboration} {STAR}),\ }\bibfield
  {title} {\bibinfo {title} {{Observation of the antimatter hypernucleus
  ${}_{\bar{{\boldsymbol{\Lambda }}}}{}^{{\bf{4}}}\bar{{\bf{H}}}$}},\ }\href
  {https://doi.org/10.1038/s41586-024-07823-0} {\bibfield  {journal} {\bibinfo
  {journal} {Nature}\ }\textbf {\bibinfo {volume} {632}},\ \bibinfo {pages}
  {1026} (\bibinfo {year} {2024})}\BibitemShut {NoStop}%
\bibitem [{\citenamefont {Acharya}\ \emph
  {et~al.}(2025{\natexlab{a}})\citenamefont {Acharya} \emph
  {et~al.}}]{ALICE:2024djx}%
  \BibitemOpen
  \bibfield  {author} {\bibinfo {author} {\bibfnamefont {S.}~\bibnamefont
  {Acharya}} \emph {et~al.} (\bibinfo {collaboration} {ALICE}),\ }\bibfield
  {title} {\bibinfo {title} {{First Measurement of $A$ = 4 Hypernuclei and
  Antihypernuclei at the LHC}},\ }\href
  {https://doi.org/10.1103/PhysRevLett.134.162301} {\bibfield  {journal}
  {\bibinfo  {journal} {Phys. Rev. Lett.}\ }\textbf {\bibinfo {volume} {134}},\
  \bibinfo {pages} {162301} (\bibinfo {year} {2025}{\natexlab{a}})}\BibitemShut
  {NoStop}%
\bibitem [{\citenamefont {{ALICE Collaboration}}(2015)}]{ALICE:2015rey}%
  \BibitemOpen
  \bibfield  {author} {\bibinfo {author} {\bibnamefont {{ALICE
  Collaboration}}},\ }\bibfield  {title} {\bibinfo {title} {{Precision
  measurement of the mass difference between light nuclei and anti-nuclei}},\
  }\href {https://doi.org/10.1038/nphys3432} {\bibfield  {journal} {\bibinfo
  {journal} {Nature Phys.}\ }\textbf {\bibinfo {volume} {11}},\ \bibinfo
  {pages} {811} (\bibinfo {year} {2015})}\BibitemShut {NoStop}%
\bibitem [{\citenamefont {Everett}\ \emph {et~al.}(2022)\citenamefont {Everett}
  \emph {et~al.}}]{JETSCAPE:2022cob}%
  \BibitemOpen
  \bibfield  {author} {\bibinfo {author} {\bibfnamefont {D.}~\bibnamefont
  {Everett}} \emph {et~al.} (\bibinfo {collaboration} {JETSCAPE}),\ }\bibfield
  {title} {\bibinfo {title} {{Role of bulk viscosity in deuteron production in
  ultrarelativistic nuclear collisions}},\ }\href
  {https://doi.org/10.1103/PhysRevC.106.064901} {\bibfield  {journal} {\bibinfo
   {journal} {Phys. Rev. C}\ }\textbf {\bibinfo {volume} {106}},\ \bibinfo
  {pages} {064901} (\bibinfo {year} {2022})}\BibitemShut {NoStop}%
\bibitem [{\citenamefont {von Doetinchem~{\it et.
  al}}(2020)}]{vonDoetinchem:2020vbj}%
  \BibitemOpen
  \bibfield  {author} {\bibinfo {author} {\bibfnamefont {P.}~\bibnamefont {von
  Doetinchem~{\it et. al}}},\ }\bibfield  {title} {\bibinfo {title}
  {{Cosmic-ray antinuclei as messengers of new physics: status and outlook for
  the new decade}},\ }\href {https://doi.org/10.1088/1475-7516/2020/08/035}
  {\bibfield  {journal} {\bibinfo  {journal} {J. Cosmol. Astropart. Phys.}\
  }\textbf {\bibinfo {volume} {08}},\ \bibinfo {pages} {035}}\BibitemShut
  {NoStop}%
\bibitem [{\citenamefont {Acharya}\ \emph
  {et~al.}(2023{\natexlab{a}})\citenamefont {Acharya} \emph
  {et~al.}}]{ALICE:2022zuz}%
  \BibitemOpen
  \bibfield  {author} {\bibinfo {author} {\bibfnamefont {S.}~\bibnamefont
  {Acharya}} \emph {et~al.} (\bibinfo {collaboration} {ALICE}),\ }\bibfield
  {title} {\bibinfo {title} {{Measurement of anti-$^{3}$He nuclei absorption in
  matter and impact on their propagation in the Galaxy}},\ }\href
  {https://doi.org/10.1038/s41567-022-01804-8} {\bibfield  {journal} {\bibinfo
  {journal} {Nature Phys.}\ }\textbf {\bibinfo {volume} {19}},\ \bibinfo
  {pages} {61} (\bibinfo {year} {2023}{\natexlab{a}})}\BibitemShut {NoStop}%
\bibitem [{\citenamefont {Butler}\ and\ \citenamefont
  {Pearson}(1963)}]{Butler:1963pp}%
  \BibitemOpen
  \bibfield  {author} {\bibinfo {author} {\bibfnamefont {S.~T.}\ \bibnamefont
  {Butler}}\ and\ \bibinfo {author} {\bibfnamefont {C.~A.}\ \bibnamefont
  {Pearson}},\ }\bibfield  {title} {\bibinfo {title} {{Deuterons from
  High-Energy Proton Bombardment of Matter}},\ }\href
  {https://doi.org/10.1103/PhysRev.129.836} {\bibfield  {journal} {\bibinfo
  {journal} {Phys. Rev.}\ }\textbf {\bibinfo {volume} {129}},\ \bibinfo {pages}
  {836} (\bibinfo {year} {1963})}\BibitemShut {NoStop}%
\bibitem [{\citenamefont {Kapusta}(1980)}]{Kapusta:1980zz}%
  \BibitemOpen
  \bibfield  {author} {\bibinfo {author} {\bibfnamefont {J.~I.}\ \bibnamefont
  {Kapusta}},\ }\bibfield  {title} {\bibinfo {title} {{Mechanisms for deuteron
  production in relativistic nuclear collisions}},\ }\href
  {https://doi.org/10.1103/PhysRevC.21.1301} {\bibfield  {journal} {\bibinfo
  {journal} {Phys. Rev. C}\ }\textbf {\bibinfo {volume} {21}},\ \bibinfo
  {pages} {1301} (\bibinfo {year} {1980})}\BibitemShut {NoStop}%
\bibitem [{\citenamefont {Csernai}\ and\ \citenamefont
  {Kapusta}(1986)}]{Csernai:1986qf}%
  \BibitemOpen
  \bibfield  {author} {\bibinfo {author} {\bibfnamefont {L.~P.}\ \bibnamefont
  {Csernai}}\ and\ \bibinfo {author} {\bibfnamefont {J.~I.}\ \bibnamefont
  {Kapusta}},\ }\bibfield  {title} {\bibinfo {title} {{Entropy and Cluster
  Production in Nuclear Collisions}},\ }\href
  {https://doi.org/10.1016/0370-1573(86)90031-1} {\bibfield  {journal}
  {\bibinfo  {journal} {Phys. Rept.}\ }\textbf {\bibinfo {volume} {131}},\
  \bibinfo {pages} {223} (\bibinfo {year} {1986})}\BibitemShut {NoStop}%
\bibitem [{\citenamefont {Braun-Munzinger}\ and\ \citenamefont
  {D\"onigus}(2019)}]{Braun-Munzinger:2018hat}%
  \BibitemOpen
  \bibfield  {author} {\bibinfo {author} {\bibfnamefont {P.}~\bibnamefont
  {Braun-Munzinger}}\ and\ \bibinfo {author} {\bibfnamefont {B.}~\bibnamefont
  {D\"onigus}},\ }\bibfield  {title} {\bibinfo {title} {{Loosely-bound objects
  produced in nuclear collisions at the LHC}},\ }\href
  {https://doi.org/10.1016/j.nuclphysa.2019.02.006} {\bibfield  {journal}
  {\bibinfo  {journal} {Nucl. Phys. A}\ }\textbf {\bibinfo {volume} {987}},\
  \bibinfo {pages} {144} (\bibinfo {year} {2019})}\BibitemShut {NoStop}%
\bibitem [{\citenamefont {Chen}\ \emph {et~al.}(2018)\citenamefont {Chen},
  \citenamefont {Keane}, \citenamefont {Ma}, \citenamefont {Tang},\ and\
  \citenamefont {Xu}}]{Chen:2018tnh}%
  \BibitemOpen
  \bibfield  {author} {\bibinfo {author} {\bibfnamefont {J.}~\bibnamefont
  {Chen}}, \bibinfo {author} {\bibfnamefont {D.}~\bibnamefont {Keane}},
  \bibinfo {author} {\bibfnamefont {Y.-G.}\ \bibnamefont {Ma}}, \bibinfo
  {author} {\bibfnamefont {A.}~\bibnamefont {Tang}},\ and\ \bibinfo {author}
  {\bibfnamefont {Z.}~\bibnamefont {Xu}},\ }\bibfield  {title} {\bibinfo
  {title} {{Antinuclei in Heavy-Ion Collisions}},\ }\href
  {https://doi.org/10.1016/j.physrep.2018.07.002} {\bibfield  {journal}
  {\bibinfo  {journal} {Phys. Rept.}\ }\textbf {\bibinfo {volume} {760}},\
  \bibinfo {pages} {1} (\bibinfo {year} {2018})}\BibitemShut {NoStop}%
\bibitem [{\citenamefont {Ono}(2019)}]{Ono:2019jxm}%
  \BibitemOpen
  \bibfield  {author} {\bibinfo {author} {\bibfnamefont {A.}~\bibnamefont
  {Ono}},\ }\bibfield  {title} {\bibinfo {title} {{Dynamics of clusters and
  fragments in heavy-ion collisions}},\ }\href
  {https://doi.org/10.1016/j.ppnp.2018.11.001} {\bibfield  {journal} {\bibinfo
  {journal} {Prog. Part. Nucl. Phys.}\ }\textbf {\bibinfo {volume} {105}},\
  \bibinfo {pages} {139} (\bibinfo {year} {2019})}\BibitemShut {NoStop}%
\bibitem [{\citenamefont {Braaten}\ \emph {et~al.}(2025)\citenamefont
  {Braaten}, \citenamefont {Ingles},\ and\ \citenamefont
  {Pickett}}]{Braaten:2024cke}%
  \BibitemOpen
  \bibfield  {author} {\bibinfo {author} {\bibfnamefont {E.}~\bibnamefont
  {Braaten}}, \bibinfo {author} {\bibfnamefont {K.}~\bibnamefont {Ingles}},\
  and\ \bibinfo {author} {\bibfnamefont {J.}~\bibnamefont {Pickett}},\
  }\bibfield  {title} {\bibinfo {title} {{Explaining Snowball-in-Hell Phenomena
  in Heavy-Ion Collisions Using a Novel Thermodynamic Variable}},\ }\href
  {https://doi.org/10.1103/7cjr-fgdw} {\bibfield  {journal} {\bibinfo
  {journal} {Phys. Rev. Lett.}\ }\textbf {\bibinfo {volume} {134}},\ \bibinfo
  {pages} {252301} (\bibinfo {year} {2025})}\BibitemShut {NoStop}%
\bibitem [{\citenamefont {Andronic}\ \emph {et~al.}(2018)\citenamefont
  {Andronic}, \citenamefont {Braun-Munzinger}, \citenamefont {Redlich},\ and\
  \citenamefont {Stachel}}]{Andronic:2017pug}%
  \BibitemOpen
  \bibfield  {author} {\bibinfo {author} {\bibfnamefont {A.}~\bibnamefont
  {Andronic}}, \bibinfo {author} {\bibfnamefont {P.}~\bibnamefont
  {Braun-Munzinger}}, \bibinfo {author} {\bibfnamefont {K.}~\bibnamefont
  {Redlich}},\ and\ \bibinfo {author} {\bibfnamefont {J.}~\bibnamefont
  {Stachel}},\ }\bibfield  {title} {\bibinfo {title} {{Decoding the phase
  structure of QCD via particle production at high energy}},\ }\href
  {https://doi.org/10.1038/s41586-018-0491-6} {\bibfield  {journal} {\bibinfo
  {journal} {Nature}\ }\textbf {\bibinfo {volume} {561}},\ \bibinfo {pages}
  {321} (\bibinfo {year} {2018})}\BibitemShut {NoStop}%
\bibitem [{\citenamefont {Scheibl}\ and\ \citenamefont
  {Heinz}(1999)}]{Scheibl:1998tk}%
  \BibitemOpen
  \bibfield  {author} {\bibinfo {author} {\bibfnamefont {R.}~\bibnamefont
  {Scheibl}}\ and\ \bibinfo {author} {\bibfnamefont {U.}~\bibnamefont
  {Heinz}},\ }\bibfield  {title} {\bibinfo {title} {Coalescence and flow in
  ultrarelativistic heavy ion collisions},\ }\href
  {https://doi.org/10.1103/PhysRevC.59.1585} {\bibfield  {journal} {\bibinfo
  {journal} {Phys. Rev. C}\ }\textbf {\bibinfo {volume} {59}},\ \bibinfo
  {pages} {1585} (\bibinfo {year} {1999})}\BibitemShut {NoStop}%
\bibitem [{\citenamefont {Bellini}\ \emph {et~al.}(2021)\citenamefont
  {Bellini}, \citenamefont {Blum}, \citenamefont {Kalweit},\ and\ \citenamefont
  {Puccio}}]{Bellini:2020cbj}%
  \BibitemOpen
  \bibfield  {author} {\bibinfo {author} {\bibfnamefont {F.}~\bibnamefont
  {Bellini}}, \bibinfo {author} {\bibfnamefont {K.}~\bibnamefont {Blum}},
  \bibinfo {author} {\bibfnamefont {A.~P.}\ \bibnamefont {Kalweit}},\ and\
  \bibinfo {author} {\bibfnamefont {M.}~\bibnamefont {Puccio}},\ }\bibfield
  {title} {\bibinfo {title} {{Examination of coalescence as the origin of
  nuclei in hadronic collisions}},\ }\href
  {https://doi.org/10.1103/PhysRevC.103.014907} {\bibfield  {journal} {\bibinfo
   {journal} {Phys. Rev. C}\ }\textbf {\bibinfo {volume} {103}},\ \bibinfo
  {pages} {014907} (\bibinfo {year} {2021})}\BibitemShut {NoStop}%
\bibitem [{\citenamefont {Oliinychenko}\ \emph {et~al.}(2019)\citenamefont
  {Oliinychenko}, \citenamefont {Pang}, \citenamefont {Elfner},\ and\
  \citenamefont {Koch}}]{Oliinychenko:2018ugs}%
  \BibitemOpen
  \bibfield  {author} {\bibinfo {author} {\bibfnamefont {D.}~\bibnamefont
  {Oliinychenko}}, \bibinfo {author} {\bibfnamefont {L.-G.}\ \bibnamefont
  {Pang}}, \bibinfo {author} {\bibfnamefont {H.}~\bibnamefont {Elfner}},\ and\
  \bibinfo {author} {\bibfnamefont {V.}~\bibnamefont {Koch}} (\bibinfo
  {collaboration} {SMASH}),\ }\bibfield  {title} {\bibinfo {title}
  {{Microscopic study of deuteron production in PbPb collisions at $\sqrt{s} =
  2.76 TeV$ via hydrodynamics and a hadronic afterburner}},\ }\href
  {https://doi.org/10.1103/PhysRevC.99.044907} {\bibfield  {journal} {\bibinfo
  {journal} {Phys. Rev. C}\ }\textbf {\bibinfo {volume} {99}},\ \bibinfo
  {pages} {044907} (\bibinfo {year} {2019})}\BibitemShut {NoStop}%
\bibitem [{\citenamefont {Sun}\ \emph {et~al.}(2024)\citenamefont {Sun},
  \citenamefont {Wang}, \citenamefont {Ko}, \citenamefont {Ma},\ and\
  \citenamefont {Shen}}]{Sun:2022xjr}%
  \BibitemOpen
  \bibfield  {author} {\bibinfo {author} {\bibfnamefont {K.-J.}\ \bibnamefont
  {Sun}}, \bibinfo {author} {\bibfnamefont {R.}~\bibnamefont {Wang}}, \bibinfo
  {author} {\bibfnamefont {C.~M.}\ \bibnamefont {Ko}}, \bibinfo {author}
  {\bibfnamefont {Y.-G.}\ \bibnamefont {Ma}},\ and\ \bibinfo {author}
  {\bibfnamefont {C.}~\bibnamefont {Shen}},\ }\bibfield  {title} {\bibinfo
  {title} {{Unveiling the dynamics of little-bang nucleosynthesis}},\ }\href
  {https://doi.org/10.1038/s41467-024-45474-x} {\bibfield  {journal} {\bibinfo
  {journal} {Nature Commun.}\ }\textbf {\bibinfo {volume} {15}},\ \bibinfo
  {pages} {1074} (\bibinfo {year} {2024})}\BibitemShut {NoStop}%
\bibitem [{\citenamefont {Adler}\ \emph {et~al.}(2005)\citenamefont {Adler}
  \emph {et~al.}}]{PHENIX:2004vqi}%
  \BibitemOpen
  \bibfield  {author} {\bibinfo {author} {\bibfnamefont {S.~S.}\ \bibnamefont
  {Adler}} \emph {et~al.} (\bibinfo {collaboration} {PHENIX}),\ }\bibfield
  {title} {\bibinfo {title} {{Deuteron and antideuteron production in Au + Au
  collisions at s(NN)**(1/2) = 200-GeV}},\ }\href
  {https://doi.org/10.1103/PhysRevLett.94.122302} {\bibfield  {journal}
  {\bibinfo  {journal} {Phys. Rev. Lett.}\ }\textbf {\bibinfo {volume} {94}},\
  \bibinfo {pages} {122302} (\bibinfo {year} {2005})}\BibitemShut {NoStop}%
\bibitem [{\citenamefont {Acharya}\ \emph {et~al.}(2022)\citenamefont {Acharya}
  \emph {et~al.}}]{ALICE:2021mfm}%
  \BibitemOpen
  \bibfield  {author} {\bibinfo {author} {\bibfnamefont {S.}~\bibnamefont
  {Acharya}} \emph {et~al.} (\bibinfo {collaboration} {ALICE}),\ }\bibfield
  {title} {\bibinfo {title} {{Production of light (anti)nuclei in pp collisions
  at $ \sqrt{s} $ = 13 TeV}},\ }\href {https://doi.org/10.1007/JHEP01(2022)106}
  {\bibfield  {journal} {\bibinfo  {journal} {JHEP}\ }\textbf {\bibinfo
  {volume} {01}},\ \bibinfo {pages} {106}}\BibitemShut {NoStop}%
\bibitem [{\citenamefont {Acharya}\ \emph
  {et~al.}(2023{\natexlab{b}})\citenamefont {Acharya} \emph
  {et~al.}}]{ALICE:2022ugx}%
  \BibitemOpen
  \bibfield  {author} {\bibinfo {author} {\bibfnamefont {S.}~\bibnamefont
  {Acharya}} \emph {et~al.} (\bibinfo {collaboration} {ALICE}),\ }\bibfield
  {title} {\bibinfo {title} {{Enhanced Deuteron Coalescence Probability in
  Jets}},\ }\href {https://doi.org/10.1103/PhysRevLett.131.042301} {\bibfield
  {journal} {\bibinfo  {journal} {Phys. Rev. Lett.}\ }\textbf {\bibinfo
  {volume} {131}},\ \bibinfo {pages} {042301} (\bibinfo {year}
  {2023}{\natexlab{b}})}\BibitemShut {NoStop}%
\bibitem [{\citenamefont {Abdulhamid}\ \emph {et~al.}(2023)\citenamefont
  {Abdulhamid} \emph {et~al.}}]{STAR:2022hbp}%
  \BibitemOpen
  \bibfield  {author} {\bibinfo {author} {\bibfnamefont {M.}~\bibnamefont
  {Abdulhamid}} \emph {et~al.} (\bibinfo {collaboration} {STAR}),\ }\bibfield
  {title} {\bibinfo {title} {{Beam Energy Dependence of Triton Production and
  Yield Ratio ($\mathrm{N}_t \times \mathrm{N}_p/\mathrm{N}_d^2$) in Au+Au
  Collisions at RHIC}},\ }\href
  {https://doi.org/10.1103/PhysRevLett.130.202301} {\bibfield  {journal}
  {\bibinfo  {journal} {Phys. Rev. Lett.}\ }\textbf {\bibinfo {volume} {130}},\
  \bibinfo {pages} {202301} (\bibinfo {year} {2023})}\BibitemShut {NoStop}%
\bibitem [{\citenamefont {Aboona}\ \emph {et~al.}(2023)\citenamefont {Aboona}
  \emph {et~al.}}]{STAR:2022fnj}%
  \BibitemOpen
  \bibfield  {author} {\bibinfo {author} {\bibfnamefont {B.}~\bibnamefont
  {Aboona}} \emph {et~al.} (\bibinfo {collaboration} {STAR}),\ }\bibfield
  {title} {\bibinfo {title} {{Observation of Directed Flow of Hypernuclei
  H\ensuremath{\Lambda}3 and H\ensuremath{\Lambda}4 in sNN=3\,\,GeV Au+Au
  Collisions at RHIC}},\ }\href
  {https://doi.org/10.1103/PhysRevLett.130.212301} {\bibfield  {journal}
  {\bibinfo  {journal} {Phys. Rev. Lett.}\ }\textbf {\bibinfo {volume} {130}},\
  \bibinfo {pages} {212301} (\bibinfo {year} {2023})}\BibitemShut {NoStop}%
\bibitem [{\citenamefont {Acharya}\ \emph
  {et~al.}(2020{\natexlab{a}})\citenamefont {Acharya} \emph
  {et~al.}}]{ALICE:2020chv}%
  \BibitemOpen
  \bibfield  {author} {\bibinfo {author} {\bibfnamefont {S.}~\bibnamefont
  {Acharya}} \emph {et~al.} (\bibinfo {collaboration} {ALICE}),\ }\bibfield
  {title} {\bibinfo {title} {{Elliptic and triangular flow of (anti)deuterons
  in Pb-Pb collisions at $\sqrt{s_{\mathrm{NN}}}$ = 5.02 TeV}},\ }\href
  {https://doi.org/10.1103/PhysRevC.102.055203} {\bibfield  {journal} {\bibinfo
   {journal} {Phys. Rev. C}\ }\textbf {\bibinfo {volume} {102}},\ \bibinfo
  {pages} {055203} (\bibinfo {year} {2020}{\natexlab{a}})}\BibitemShut
  {NoStop}%
\bibitem [{\citenamefont {Lednicky}(2002)}]{Lednicky:2002fq}%
  \BibitemOpen
  \bibfield  {author} {\bibinfo {author} {\bibfnamefont {R.}~\bibnamefont
  {Lednicky}},\ }\bibfield  {title} {\bibinfo {title} {{Progress in correlation
  femtoscopy}},\ }in\ \href {https://doi.org/10.1142/9789812704962_0005} {\emph
  {\bibinfo {booktitle} {{32nd International Symposium on Multiparticle
  Dynamics}}}}\ (\bibinfo {year} {2002})\ pp.\ \bibinfo {pages} {21--26},\
  \Eprint {https://arxiv.org/abs/nucl-th/0212089} {arXiv:nucl-th/0212089}
  \BibitemShut {NoStop}%
\bibitem [{\citenamefont {Lisa}\ \emph {et~al.}(2005)\citenamefont {Lisa},
  \citenamefont {Pratt}, \citenamefont {Soltz},\ and\ \citenamefont
  {Wiedemann}}]{Lisa:2005dd}%
  \BibitemOpen
  \bibfield  {author} {\bibinfo {author} {\bibfnamefont {M.~A.}\ \bibnamefont
  {Lisa}}, \bibinfo {author} {\bibfnamefont {S.}~\bibnamefont {Pratt}},
  \bibinfo {author} {\bibfnamefont {R.}~\bibnamefont {Soltz}},\ and\ \bibinfo
  {author} {\bibfnamefont {U.}~\bibnamefont {Wiedemann}},\ }\bibfield  {title}
  {\bibinfo {title} {{Femtoscopy in relativistic heavy ion collisions}},\
  }\href {https://doi.org/10.1146/annurev.nucl.55.090704.151533} {\bibfield
  {journal} {\bibinfo  {journal} {Ann. Rev. Nucl. Part. Sci.}\ }\textbf
  {\bibinfo {volume} {55}},\ \bibinfo {pages} {357} (\bibinfo {year}
  {2005})}\BibitemShut {NoStop}%
\bibitem [{\citenamefont {Hanbury~Brown}\ and\ \citenamefont
  {Twiss}(1956)}]{HanburyBrown:1956bqd}%
  \BibitemOpen
  \bibfield  {author} {\bibinfo {author} {\bibfnamefont {R.}~\bibnamefont
  {Hanbury~Brown}}\ and\ \bibinfo {author} {\bibfnamefont {R.~Q.}\ \bibnamefont
  {Twiss}},\ }\bibfield  {title} {\bibinfo {title} {{A Test of a new type of
  stellar interferometer on Sirius}},\ }\href
  {https://doi.org/10.1038/1781046a0} {\bibfield  {journal} {\bibinfo
  {journal} {Nature}\ }\textbf {\bibinfo {volume} {178}},\ \bibinfo {pages}
  {1046} (\bibinfo {year} {1956})}\BibitemShut {NoStop}%
\bibitem [{\citenamefont {Baym}(1998)}]{Baym:1997ce}%
  \BibitemOpen
  \bibfield  {author} {\bibinfo {author} {\bibfnamefont {G.}~\bibnamefont
  {Baym}},\ }\bibfield  {title} {\bibinfo {title} {{The Physics of Hanbury
  Brown-Twiss intensity interferometry: From stars to nuclear collisions}},\
  }\href@noop {} {\bibfield  {journal} {\bibinfo  {journal} {Acta Phys. Polon.
  B}\ }\textbf {\bibinfo {volume} {29}},\ \bibinfo {pages} {1839} (\bibinfo
  {year} {1998})}\BibitemShut {NoStop}%
\bibitem [{\citenamefont {Heinz}\ and\ \citenamefont
  {Jacak}(1999)}]{Heinz:1999rw}%
  \BibitemOpen
  \bibfield  {author} {\bibinfo {author} {\bibfnamefont {U.~W.}\ \bibnamefont
  {Heinz}}\ and\ \bibinfo {author} {\bibfnamefont {B.~V.}\ \bibnamefont
  {Jacak}},\ }\bibfield  {title} {\bibinfo {title} {{Two particle correlations
  in relativistic heavy ion collisions}},\ }\href
  {https://doi.org/10.1146/annurev.nucl.49.1.529} {\bibfield  {journal}
  {\bibinfo  {journal} {Ann. Rev. Nucl. Part. Sci.}\ }\textbf {\bibinfo
  {volume} {49}},\ \bibinfo {pages} {529} (\bibinfo {year} {1999})}\BibitemShut
  {NoStop}%
\bibitem [{\citenamefont {Lednicky}\ and\ \citenamefont
  {Lyuboshits}(1981)}]{Lednicky:1981su}%
  \BibitemOpen
  \bibfield  {author} {\bibinfo {author} {\bibfnamefont {R.}~\bibnamefont
  {Lednicky}}\ and\ \bibinfo {author} {\bibfnamefont {V.~L.}\ \bibnamefont
  {Lyuboshits}},\ }\bibfield  {title} {\bibinfo {title} {{Final State
  Interaction Effect on Pairing Correlations Between Particles with Small
  Relative Momenta}},\ }\href@noop {} {\bibfield  {journal} {\bibinfo
  {journal} {Yad. Fiz.}\ }\textbf {\bibinfo {volume} {35}},\ \bibinfo {pages}
  {1316} (\bibinfo {year} {1981})}\BibitemShut {NoStop}%
\bibitem [{\citenamefont {Adamczyk}\ \emph {et~al.}(2015)\citenamefont
  {Adamczyk} \emph {et~al.}}]{STAR:2015kha}%
  \BibitemOpen
  \bibfield  {author} {\bibinfo {author} {\bibfnamefont {L.}~\bibnamefont
  {Adamczyk}} \emph {et~al.} (\bibinfo {collaboration} {STAR}),\ }\bibfield
  {title} {\bibinfo {title} {{Measurement of Interaction between
  Antiprotons}},\ }\href {https://doi.org/10.1038/nature15724} {\bibfield
  {journal} {\bibinfo  {journal} {Nature}\ }\textbf {\bibinfo {volume} {527}},\
  \bibinfo {pages} {345} (\bibinfo {year} {2015})}\BibitemShut {NoStop}%
\bibitem [{\citenamefont {Collaboration}\ \emph {et~al.}(2020)\citenamefont
  {Collaboration} \emph {et~al.}}]{ALICE:2020mfd}%
  \BibitemOpen
  \bibfield  {author} {\bibinfo {author} {\bibfnamefont {A.}~\bibnamefont
  {Collaboration}} \emph {et~al.} (\bibinfo {collaboration} {ALICE}),\
  }\bibfield  {title} {\bibinfo {title} {{Unveiling the strong interaction
  among hadrons at the LHC}},\ }\href
  {https://doi.org/10.1038/s41586-020-3001-6} {\bibfield  {journal} {\bibinfo
  {journal} {Nature}\ }\textbf {\bibinfo {volume} {588}},\ \bibinfo {pages}
  {232} (\bibinfo {year} {2020})},\ \bibinfo {note} {[Erratum: Nature 590, E13
  (2021)]}\BibitemShut {NoStop}%
\bibitem [{\citenamefont {Acharya}\ \emph
  {et~al.}(2020{\natexlab{b}})\citenamefont {Acharya} \emph
  {et~al.}}]{ALICE:2019gcn}%
  \BibitemOpen
  \bibfield  {author} {\bibinfo {author} {\bibfnamefont {S.}~\bibnamefont
  {Acharya}} \emph {et~al.} (\bibinfo {collaboration} {ALICE}),\ }\bibfield
  {title} {\bibinfo {title} {{Scattering studies with low-energy kaon-proton
  femtoscopy in proton-proton collisions at the LHC}},\ }\href
  {https://doi.org/10.1103/PhysRevLett.124.092301} {\bibfield  {journal}
  {\bibinfo  {journal} {Phys. Rev. Lett.}\ }\textbf {\bibinfo {volume} {124}},\
  \bibinfo {pages} {092301} (\bibinfo {year} {2020}{\natexlab{b}})}\BibitemShut
  {NoStop}%
\bibitem [{\citenamefont {Si}\ \emph {et~al.}(2025)\citenamefont {Si} \emph
  {et~al.}}]{Si:2025eou}%
  \BibitemOpen
  \bibfield  {author} {\bibinfo {author} {\bibfnamefont {D.}~\bibnamefont {Si}}
  \emph {et~al.},\ }\bibfield  {title} {\bibinfo {title} {{Extracting
  Neutron-Neutron Interaction Strength and Spatiotemporal Dynamics of Neutron
  Emission from the Two-Particle Correlation Function}},\ }\href
  {https://doi.org/10.1103/PhysRevLett.134.222301} {\bibfield  {journal}
  {\bibinfo  {journal} {Phys. Rev. Lett.}\ }\textbf {\bibinfo {volume} {134}},\
  \bibinfo {pages} {222301} (\bibinfo {year} {2025})}\BibitemShut {NoStop}%
\bibitem [{\citenamefont {Liu}\ \emph {et~al.}(2025{\natexlab{a}})\citenamefont
  {Liu}, \citenamefont {Pan}, \citenamefont {Liu}, \citenamefont {Wu},
  \citenamefont {Lu},\ and\ \citenamefont {Geng}}]{Liu:2024uxn}%
  \BibitemOpen
  \bibfield  {author} {\bibinfo {author} {\bibfnamefont {M.-Z.}\ \bibnamefont
  {Liu}}, \bibinfo {author} {\bibfnamefont {Y.-W.}\ \bibnamefont {Pan}},
  \bibinfo {author} {\bibfnamefont {Z.-W.}\ \bibnamefont {Liu}}, \bibinfo
  {author} {\bibfnamefont {T.-W.}\ \bibnamefont {Wu}}, \bibinfo {author}
  {\bibfnamefont {J.-X.}\ \bibnamefont {Lu}},\ and\ \bibinfo {author}
  {\bibfnamefont {L.-S.}\ \bibnamefont {Geng}},\ }\bibfield  {title} {\bibinfo
  {title} {{Three ways to decipher the nature of exotic hadrons: Multiplets,
  three-body hadronic molecules, and correlation functions}},\ }\href
  {https://doi.org/10.1016/j.physrep.2024.12.001} {\bibfield  {journal}
  {\bibinfo  {journal} {Phys. Rept.}\ }\textbf {\bibinfo {volume} {1108}},\
  \bibinfo {pages} {1} (\bibinfo {year} {2025}{\natexlab{a}})}\BibitemShut
  {NoStop}%
\bibitem [{\citenamefont {Koonin}(1977)}]{Koonin:1977fh}%
  \BibitemOpen
  \bibfield  {author} {\bibinfo {author} {\bibfnamefont {S.~E.}\ \bibnamefont
  {Koonin}},\ }\bibfield  {title} {\bibinfo {title} {{Proton Pictures of
  High-Energy Nuclear Collisions}},\ }\href
  {https://doi.org/10.1016/0370-2693(77)90340-9} {\bibfield  {journal}
  {\bibinfo  {journal} {Phys. Lett. B}\ }\textbf {\bibinfo {volume} {70}},\
  \bibinfo {pages} {43} (\bibinfo {year} {1977})}\BibitemShut {NoStop}%
\bibitem [{\citenamefont {Wiedemann}\ and\ \citenamefont
  {Heinz}(1997)}]{Wiedemann:1996ig}%
  \BibitemOpen
  \bibfield  {author} {\bibinfo {author} {\bibfnamefont {U.~A.}\ \bibnamefont
  {Wiedemann}}\ and\ \bibinfo {author} {\bibfnamefont {U.~W.}\ \bibnamefont
  {Heinz}},\ }\bibfield  {title} {\bibinfo {title} {{Resonance contributions to
  HBT correlation radii}},\ }\href {https://doi.org/10.1103/PhysRevC.56.3265}
  {\bibfield  {journal} {\bibinfo  {journal} {Phys. Rev. C}\ }\textbf {\bibinfo
  {volume} {56}},\ \bibinfo {pages} {3265} (\bibinfo {year}
  {1997})}\BibitemShut {NoStop}%
\bibitem [{\citenamefont {Fabbietti}\ \emph {et~al.}(2021)\citenamefont
  {Fabbietti}, \citenamefont {Mantovani~Sarti},\ and\ \citenamefont
  {Vazquez~Doce}}]{Fabbietti:2020bfg}%
  \BibitemOpen
  \bibfield  {author} {\bibinfo {author} {\bibfnamefont {L.}~\bibnamefont
  {Fabbietti}}, \bibinfo {author} {\bibfnamefont {V.}~\bibnamefont
  {Mantovani~Sarti}},\ and\ \bibinfo {author} {\bibfnamefont {O.}~\bibnamefont
  {Vazquez~Doce}},\ }\bibfield  {title} {\bibinfo {title} {{Study of the Strong
  Interaction Among Hadrons with Correlations at the LHC}},\ }\href
  {https://doi.org/10.1146/annurev-nucl-102419-034438} {\bibfield  {journal}
  {\bibinfo  {journal} {Ann. Rev. Nucl. Part. Sci.}\ }\textbf {\bibinfo
  {volume} {71}},\ \bibinfo {pages} {377} (\bibinfo {year} {2021})}\BibitemShut
  {NoStop}%
\bibitem [{\citenamefont {Acharya}\ \emph {et~al.}(2024)\citenamefont {Acharya}
  \emph {et~al.}}]{ALICE:2023bny}%
  \BibitemOpen
  \bibfield  {author} {\bibinfo {author} {\bibfnamefont {S.}~\bibnamefont
  {Acharya}} \emph {et~al.} (\bibinfo {collaboration} {ALICE}),\ }\bibfield
  {title} {\bibinfo {title} {{Exploring the Strong Interaction of Three-Body
  Systems at the LHC}},\ }\href {https://doi.org/10.1103/PhysRevX.14.031051}
  {\bibfield  {journal} {\bibinfo  {journal} {Phys. Rev. X}\ }\textbf {\bibinfo
  {volume} {14}},\ \bibinfo {pages} {031051} (\bibinfo {year}
  {2024})}\BibitemShut {NoStop}%
\bibitem [{\citenamefont {Mrowczynski}(2025)}]{Mrowczynski:2025qys}%
  \BibitemOpen
  \bibfield  {author} {\bibinfo {author} {\bibfnamefont {S.}~\bibnamefont
  {Mrowczynski}},\ }\bibfield  {title} {\bibinfo {title} {{Two- versus
  three-body approach to femtoscopic hadron-deuteron correlations}},\ }\href
  {https://doi.org/10.1016/j.physletb.2025.139413} {\bibfield  {journal}
  {\bibinfo  {journal} {Phys. Lett. B}\ }\textbf {\bibinfo {volume} {864}},\
  \bibinfo {pages} {139413} (\bibinfo {year} {2025})}\BibitemShut {NoStop}%
\bibitem [{\citenamefont {V{\'a}zquez~Doce}\ \emph {et~al.}(2025)\citenamefont
  {V{\'a}zquez~Doce}, \citenamefont {Mihaylov},\ and\ \citenamefont
  {Fabbietti}}]{VazquezDoce:2024nye}%
  \BibitemOpen
  \bibfield  {author} {\bibinfo {author} {\bibfnamefont {O.}~\bibnamefont
  {V{\'a}zquez~Doce}}, \bibinfo {author} {\bibfnamefont {D.}~\bibnamefont
  {Mihaylov}},\ and\ \bibinfo {author} {\bibfnamefont {L.}~\bibnamefont
  {Fabbietti}},\ }\bibfield  {title} {\bibinfo {title} {{Study of the deuterons
  emission time in pp collisions at the LHC via kaon-deuteron correlations}},\
  }\href {https://doi.org/10.1140/epja/s10050-025-01534-4} {\bibfield
  {journal} {\bibinfo  {journal} {Eur. Phys. J. A}\ }\textbf {\bibinfo {volume}
  {61}},\ \bibinfo {pages} {53} (\bibinfo {year} {2025})}\BibitemShut {NoStop}%
\bibitem [{\citenamefont {Mr{\'o}wczy{\'n}ski}\ and\ \citenamefont
  {S{\l}o{\'n}}(2021)}]{Mrowczynski:2021bzy}%
  \BibitemOpen
  \bibfield  {author} {\bibinfo {author} {\bibfnamefont {S.}~\bibnamefont
  {Mr{\'o}wczy{\'n}ski}}\ and\ \bibinfo {author} {\bibfnamefont
  {P.}~\bibnamefont {S{\l}o{\'n}}},\ }\bibfield  {title} {\bibinfo {title}
  {{Deuteron-deuteron correlation function in nucleus-nucleus collisions}},\
  }\href {https://doi.org/10.1103/PhysRevC.104.024909} {\bibfield  {journal}
  {\bibinfo  {journal} {Phys. Rev. C}\ }\textbf {\bibinfo {volume} {104}},\
  \bibinfo {pages} {024909} (\bibinfo {year} {2021})}\BibitemShut {NoStop}%
\bibitem [{\citenamefont {Acharya}\ \emph
  {et~al.}(2025{\natexlab{b}})\citenamefont {Acharya} \emph
  {et~al.}}]{ALICE:2025aur}%
  \BibitemOpen
  \bibfield  {author} {\bibinfo {author} {\bibfnamefont {S.}~\bibnamefont
  {Acharya}} \emph {et~al.} (\bibinfo {collaboration} {ALICE}),\ }\bibfield
  {title} {\bibinfo {title} {{Investigating the ${\text {p--}\pi ^{\pm }}$ and
  ${\text {p--p--}\pi ^{\pm }}$ dynamics with femtoscopy in pp collisions at
  ${\sqrt{\textit{s}}=13}$~TeV}},\ }\href
  {https://doi.org/10.1140/epja/s10050-025-01615-4} {\bibfield  {journal}
  {\bibinfo  {journal} {Eur. Phys. J. A}\ }\textbf {\bibinfo {volume} {61}},\
  \bibinfo {pages} {194} (\bibinfo {year} {2025}{\natexlab{b}})}\BibitemShut
  {NoStop}%
\bibitem [{\citenamefont {Acharya}\ \emph
  {et~al.}(2025{\natexlab{c}})\citenamefont {Acharya} \emph
  {et~al.}}]{ALICE:2025byl}%
  \BibitemOpen
  \bibfield  {author} {\bibinfo {author} {\bibfnamefont {S.}~\bibnamefont
  {Acharya}} \emph {et~al.} (\bibinfo {collaboration} {ALICE}),\ }\bibfield
  {title} {\bibinfo {title} {{Observation of deuteron and antideuteron
  formation from resonance-decay nucleons}},\ }\href
  {https://doi.org/10.1038/s41586-025-09775-5} {\bibfield  {journal} {\bibinfo
  {journal} {Nature}\ }\textbf {\bibinfo {volume} {648}},\ \bibinfo {pages}
  {306} (\bibinfo {year} {2025}{\natexlab{c}})},\ \Eprint
  {https://arxiv.org/abs/2504.02393} {arXiv:2504.02393 [nucl-ex]} \BibitemShut
  {NoStop}%
\bibitem [{\citenamefont {Lin}\ \emph {et~al.}(2005)\citenamefont {Lin},
  \citenamefont {Ko}, \citenamefont {Li}, \citenamefont {Zhang},\ and\
  \citenamefont {Pal}}]{Lin:2004en}%
  \BibitemOpen
  \bibfield  {author} {\bibinfo {author} {\bibfnamefont {Z.-W.}\ \bibnamefont
  {Lin}}, \bibinfo {author} {\bibfnamefont {C.~M.}\ \bibnamefont {Ko}},
  \bibinfo {author} {\bibfnamefont {B.-A.}\ \bibnamefont {Li}}, \bibinfo
  {author} {\bibfnamefont {B.}~\bibnamefont {Zhang}},\ and\ \bibinfo {author}
  {\bibfnamefont {S.}~\bibnamefont {Pal}},\ }\bibfield  {title} {\bibinfo
  {title} {{A Multi-phase transport model for relativistic heavy ion
  collisions}},\ }\href {https://doi.org/10.1103/PhysRevC.72.064901} {\bibfield
   {journal} {\bibinfo  {journal} {Phys. Rev. C}\ }\textbf {\bibinfo {volume}
  {72}},\ \bibinfo {pages} {064901} (\bibinfo {year} {2005})}\BibitemShut
  {NoStop}%
\bibitem [{\citenamefont {Ma}\ and\ \citenamefont {Wang}(2011)}]{Ma:2010dv}%
  \BibitemOpen
  \bibfield  {author} {\bibinfo {author} {\bibfnamefont {G.-L.}\ \bibnamefont
  {Ma}}\ and\ \bibinfo {author} {\bibfnamefont {X.-N.}\ \bibnamefont {Wang}},\
  }\bibfield  {title} {\bibinfo {title} {{Jets, Mach cone, hot spots, ridges,
  harmonic flow, dihadron and $\gamma$-hadron correlation in high-energy
  heavy-ion collisions}},\ }\href
  {https://doi.org/10.1103/PhysRevLett.106.162301} {\bibfield  {journal}
  {\bibinfo  {journal} {Phys. Rev. Lett.}\ }\textbf {\bibinfo {volume} {106}},\
  \bibinfo {pages} {162301} (\bibinfo {year} {2011})},\ \Eprint
  {https://arxiv.org/abs/1011.5249} {arXiv:1011.5249 [nucl-th]} \BibitemShut
  {NoStop}%
\bibitem [{\citenamefont {Luo}\ \emph {et~al.}(2023)\citenamefont {Luo},
  \citenamefont {Mao}, \citenamefont {Qin}, \citenamefont {Wang},\ and\
  \citenamefont {Zhang}}]{Luo:2021voy}%
  \BibitemOpen
  \bibfield  {author} {\bibinfo {author} {\bibfnamefont {A.}~\bibnamefont
  {Luo}}, \bibinfo {author} {\bibfnamefont {Y.-X.}\ \bibnamefont {Mao}},
  \bibinfo {author} {\bibfnamefont {G.-Y.}\ \bibnamefont {Qin}}, \bibinfo
  {author} {\bibfnamefont {E.-K.}\ \bibnamefont {Wang}},\ and\ \bibinfo
  {author} {\bibfnamefont {H.-Z.}\ \bibnamefont {Zhang}},\ }\bibfield  {title}
  {\bibinfo {title} {{Enhancement of baryon-to-meson ratios around jets as a
  signature of medium response}},\ }\href
  {https://doi.org/10.1016/j.physletb.2022.137638} {\bibfield  {journal}
  {\bibinfo  {journal} {Phys. Lett. B}\ }\textbf {\bibinfo {volume} {837}},\
  \bibinfo {pages} {137638} (\bibinfo {year} {2023})},\ \Eprint
  {https://arxiv.org/abs/2109.14314} {arXiv:2109.14314 [hep-ph]} \BibitemShut
  {NoStop}%
\bibitem [{\citenamefont {Oh}\ \emph {et~al.}(2009)\citenamefont {Oh},
  \citenamefont {Lin},\ and\ \citenamefont {Ko}}]{Oh:2009gx}%
  \BibitemOpen
  \bibfield  {author} {\bibinfo {author} {\bibfnamefont {Y.}~\bibnamefont
  {Oh}}, \bibinfo {author} {\bibfnamefont {Z.-W.}\ \bibnamefont {Lin}},\ and\
  \bibinfo {author} {\bibfnamefont {C.~M.}\ \bibnamefont {Ko}},\ }\bibfield
  {title} {\bibinfo {title} {{Deuteron production and elliptic flow in
  relativistic heavy ion collisions}},\ }\href
  {https://doi.org/10.1103/PhysRevC.80.064902} {\bibfield  {journal} {\bibinfo
  {journal} {Phys. Rev. C}\ }\textbf {\bibinfo {volume} {80}},\ \bibinfo
  {pages} {064902} (\bibinfo {year} {2009})},\ \Eprint
  {https://arxiv.org/abs/0910.1977} {arXiv:0910.1977 [nucl-th]} \BibitemShut
  {NoStop}%
\bibitem [{\citenamefont {Shao}\ \emph {et~al.}(2022)\citenamefont {Shao},
  \citenamefont {Chen}, \citenamefont {Ma},\ and\ \citenamefont
  {Xu}}]{Shao:2022eyd}%
  \BibitemOpen
  \bibfield  {author} {\bibinfo {author} {\bibfnamefont {T.}~\bibnamefont
  {Shao}}, \bibinfo {author} {\bibfnamefont {J.}~\bibnamefont {Chen}}, \bibinfo
  {author} {\bibfnamefont {Y.-G.}\ \bibnamefont {Ma}},\ and\ \bibinfo {author}
  {\bibfnamefont {Z.}~\bibnamefont {Xu}},\ }\bibfield  {title} {\bibinfo
  {title} {{Production of light antinuclei in pp collisions by dynamical
  coalescence and their fluxes in cosmic rays near earth}},\ }\href
  {https://doi.org/10.1103/PhysRevC.105.065801} {\bibfield  {journal} {\bibinfo
   {journal} {Phys. Rev. C}\ }\textbf {\bibinfo {volume} {105}},\ \bibinfo
  {pages} {065801} (\bibinfo {year} {2022})},\ \Eprint
  {https://arxiv.org/abs/2205.13626} {arXiv:2205.13626 [hep-ph]} \BibitemShut
  {NoStop}%
\bibitem [{\citenamefont {Wang}\ \emph {et~al.}(2022)\citenamefont {Wang},
  \citenamefont {Zhang},\ and\ \citenamefont {Ma}}]{Wang:2021ghq}%
  \BibitemOpen
  \bibfield  {author} {\bibinfo {author} {\bibfnamefont {Y.-Z.}\ \bibnamefont
  {Wang}}, \bibinfo {author} {\bibfnamefont {S.}~\bibnamefont {Zhang}},\ and\
  \bibinfo {author} {\bibfnamefont {Y.-G.}\ \bibnamefont {Ma}},\ }\bibfield
  {title} {\bibinfo {title} {{System dependence of away-side broadening and
  {\ensuremath{\alpha}}-clustering light nuclei structure effect in dihadron
  azimuthal correlations}},\ }\href
  {https://doi.org/10.1016/j.physletb.2022.137198} {\bibfield  {journal}
  {\bibinfo  {journal} {Phys. Lett. B}\ }\textbf {\bibinfo {volume} {831}},\
  \bibinfo {pages} {137198} (\bibinfo {year} {2022})},\ \Eprint
  {https://arxiv.org/abs/2112.08617} {arXiv:2112.08617 [nucl-th]} \BibitemShut
  {NoStop}%
\bibitem [{\citenamefont {Danielewicz}\ and\ \citenamefont
  {Bertsch}(1991)}]{Danielewicz:1991dh}%
  \BibitemOpen
  \bibfield  {author} {\bibinfo {author} {\bibfnamefont {P.}~\bibnamefont
  {Danielewicz}}\ and\ \bibinfo {author} {\bibfnamefont {G.~F.}\ \bibnamefont
  {Bertsch}},\ }\bibfield  {title} {\bibinfo {title} {{Production of deuterons
  and pions in a transport model of energetic heavy ion reactions}},\ }\href
  {https://doi.org/10.1016/0375-9474(91)90541-D} {\bibfield  {journal}
  {\bibinfo  {journal} {Nucl. Phys. A}\ }\textbf {\bibinfo {volume} {533}},\
  \bibinfo {pages} {712} (\bibinfo {year} {1991})}\BibitemShut {NoStop}%
\bibitem [{\citenamefont {Wang}\ \emph {et~al.}(2026)\citenamefont {Wang},
  \citenamefont {Zhang}, \citenamefont {Ma}, \citenamefont {Chen},
  \citenamefont {Ko},\ and\ \citenamefont {Sun}}]{Wang:2025wim}%
  \BibitemOpen
  \bibfield  {author} {\bibinfo {author} {\bibfnamefont {R.}~\bibnamefont
  {Wang}}, \bibinfo {author} {\bibfnamefont {Z.}~\bibnamefont {Zhang}},
  \bibinfo {author} {\bibfnamefont {Y.-G.}\ \bibnamefont {Ma}}, \bibinfo
  {author} {\bibfnamefont {L.-W.}\ \bibnamefont {Chen}}, \bibinfo {author}
  {\bibfnamefont {C.~M.}\ \bibnamefont {Ko}},\ and\ \bibinfo {author}
  {\bibfnamefont {K.-J.}\ \bibnamefont {Sun}},\ }\bibfield  {title} {\bibinfo
  {title} {{Clustered Nature of Hot and Dense Nuclear Matter: Signatures from
  Heavy-Ion Collisions}},\ }\href {https://doi.org/10.1103/ffxr-mmlg}
  {\bibfield  {journal} {\bibinfo  {journal} {Phys. Rev. Lett.}\ }\textbf
  {\bibinfo {volume} {136}},\ \bibinfo {pages} {172301} (\bibinfo {year}
  {2026})},\ \Eprint {https://arxiv.org/abs/2507.16613} {arXiv:2507.16613
  [nucl-th]} \BibitemShut {NoStop}%
\bibitem [{\citenamefont {Xu}\ and\ \citenamefont {Greiner}(2005)}]{Xu:2004mz}%
  \BibitemOpen
  \bibfield  {author} {\bibinfo {author} {\bibfnamefont {Z.}~\bibnamefont
  {Xu}}\ and\ \bibinfo {author} {\bibfnamefont {C.}~\bibnamefont {Greiner}},\
  }\bibfield  {title} {\bibinfo {title} {{Thermalization of gluons in
  ultrarelativistic heavy ion collisions by including three-body interactions
  in a parton cascade}},\ }\href {https://doi.org/10.1103/PhysRevC.71.064901}
  {\bibfield  {journal} {\bibinfo  {journal} {Phys. Rev. C}\ }\textbf {\bibinfo
  {volume} {71}},\ \bibinfo {pages} {064901} (\bibinfo {year}
  {2005})}\BibitemShut {NoStop}%
\bibitem [{\citenamefont {Wang}\ \emph {et~al.}(2020)\citenamefont {Wang},
  \citenamefont {Zhang}, \citenamefont {Chen},\ and\ \citenamefont
  {Ma}}]{Wang:2020ixf}%
  \BibitemOpen
  \bibfield  {author} {\bibinfo {author} {\bibfnamefont {R.}~\bibnamefont
  {Wang}}, \bibinfo {author} {\bibfnamefont {Z.}~\bibnamefont {Zhang}},
  \bibinfo {author} {\bibfnamefont {L.-W.}\ \bibnamefont {Chen}},\ and\
  \bibinfo {author} {\bibfnamefont {Y.-G.}\ \bibnamefont {Ma}},\ }\bibfield
  {title} {\bibinfo {title} {{Nuclear Collective Dynamics in Transport Model
  With the Lattice Hamiltonian Method}},\ }\href
  {https://doi.org/10.3389/fphy.2020.00330} {\bibfield  {journal} {\bibinfo
  {journal} {Front. in Phys.}\ }\textbf {\bibinfo {volume} {8}},\ \bibinfo
  {pages} {330} (\bibinfo {year} {2020})}\BibitemShut {NoStop}%
\bibitem [{\citenamefont {Zyla}\ \emph {et~al.}(2020)\citenamefont {Zyla} \emph
  {et~al.}}]{ParticleDataGroup:2020ssz}%
  \BibitemOpen
  \bibfield  {author} {\bibinfo {author} {\bibfnamefont {P.~A.}\ \bibnamefont
  {Zyla}} \emph {et~al.} (\bibinfo {collaboration} {Particle Data Group}),\
  }\bibfield  {title} {\bibinfo {title} {{Review of Particle Physics}},\ }\href
  {https://doi.org/10.1093/ptep/ptaa104} {\bibfield  {journal} {\bibinfo
  {journal} {PTEP}\ }\textbf {\bibinfo {volume} {2020}},\ \bibinfo {pages}
  {083C01} (\bibinfo {year} {2020})}\BibitemShut {NoStop}%
\bibitem [{\citenamefont {Wong}(1982)}]{Wong:1982zzb}%
  \BibitemOpen
  \bibfield  {author} {\bibinfo {author} {\bibfnamefont {C.-Y.}\ \bibnamefont
  {Wong}},\ }\bibfield  {title} {\bibinfo {title} {{Dynamics of nuclear fluid.
  VIII. Time-dependent Hartree-Fock approximation from a classical point of
  view}},\ }\href {https://doi.org/10.1103/PhysRevC.25.1460} {\bibfield
  {journal} {\bibinfo  {journal} {Phys. Rev. C}\ }\textbf {\bibinfo {volume}
  {25}},\ \bibinfo {pages} {1460} (\bibinfo {year} {1982})}\BibitemShut
  {NoStop}%
\bibitem [{\citenamefont {van Hees}\ and\ \citenamefont
  {Rapp}(2005)}]{vanHees:2004vt}%
  \BibitemOpen
  \bibfield  {author} {\bibinfo {author} {\bibfnamefont {H.}~\bibnamefont {van
  Hees}}\ and\ \bibinfo {author} {\bibfnamefont {R.}~\bibnamefont {Rapp}},\
  }\bibfield  {title} {\bibinfo {title} {{Delta(1232) and nucleon spectral
  functions in hot hadronic matter}},\ }\href
  {https://doi.org/10.1016/j.physletb.2004.10.062} {\bibfield  {journal}
  {\bibinfo  {journal} {Phys. Lett. B}\ }\textbf {\bibinfo {volume} {606}},\
  \bibinfo {pages} {59} (\bibinfo {year} {2005})},\ \Eprint
  {https://arxiv.org/abs/nucl-th/0407050} {arXiv:nucl-th/0407050} \BibitemShut
  {NoStop}%
\bibitem [{\citenamefont {Rais}\ \emph {et~al.}(2022)\citenamefont {Rais},
  \citenamefont {van Hees},\ and\ \citenamefont {Greiner}}]{Rais:2022gfg}%
  \BibitemOpen
  \bibfield  {author} {\bibinfo {author} {\bibfnamefont {J.}~\bibnamefont
  {Rais}}, \bibinfo {author} {\bibfnamefont {H.}~\bibnamefont {van Hees}},\
  and\ \bibinfo {author} {\bibfnamefont {C.}~\bibnamefont {Greiner}},\
  }\bibfield  {title} {\bibinfo {title} {{Bound-state formation in
  time-dependent potentials}},\ }\href
  {https://doi.org/10.1103/PhysRevC.106.064004} {\bibfield  {journal} {\bibinfo
   {journal} {Phys. Rev. C}\ }\textbf {\bibinfo {volume} {106}},\ \bibinfo
  {pages} {064004} (\bibinfo {year} {2022})},\ \Eprint
  {https://arxiv.org/abs/2207.04898} {arXiv:2207.04898 [quant-ph]} \BibitemShut
  {NoStop}%
\bibitem [{\citenamefont {Mr{\'o}wczy{\'n}ski}\ and\ \citenamefont
  {S{\l}o{\'n}}(2020)}]{Mrowczynski:2019yrr}%
  \BibitemOpen
  \bibfield  {author} {\bibinfo {author} {\bibfnamefont {S.}~\bibnamefont
  {Mr{\'o}wczy{\'n}ski}}\ and\ \bibinfo {author} {\bibfnamefont
  {P.}~\bibnamefont {S{\l}o{\'n}}},\ }\bibfield  {title} {\bibinfo {title}
  {{Hadron{\textendash}Deuteron Correlations and Production of Light Nuclei in
  Relativistic Heavy-ion Collisions}},\ }\href
  {https://doi.org/10.5506/APhysPolB.51.1739} {\bibfield  {journal} {\bibinfo
  {journal} {Acta Phys. Polon. B}\ }\textbf {\bibinfo {volume} {51}},\ \bibinfo
  {pages} {1739} (\bibinfo {year} {2020})},\ \Eprint
  {https://arxiv.org/abs/1904.08320} {arXiv:1904.08320 [nucl-th]} \BibitemShut
  {NoStop}%
\bibitem [{\citenamefont {Typel}\ \emph {et~al.}(2010)\citenamefont {Typel},
  \citenamefont {Ropke}, \citenamefont {Klahn}, \citenamefont {Blaschke},\ and\
  \citenamefont {Wolter}}]{Typel:2009sy}%
  \BibitemOpen
  \bibfield  {author} {\bibinfo {author} {\bibfnamefont {S.}~\bibnamefont
  {Typel}}, \bibinfo {author} {\bibfnamefont {G.}~\bibnamefont {Ropke}},
  \bibinfo {author} {\bibfnamefont {T.}~\bibnamefont {Klahn}}, \bibinfo
  {author} {\bibfnamefont {D.}~\bibnamefont {Blaschke}},\ and\ \bibinfo
  {author} {\bibfnamefont {H.~H.}\ \bibnamefont {Wolter}},\ }\bibfield  {title}
  {\bibinfo {title} {{Composition and thermodynamics of nuclear matter with
  light clusters}},\ }\href {https://doi.org/10.1103/PhysRevC.81.015803}
  {\bibfield  {journal} {\bibinfo  {journal} {Phys. Rev. C}\ }\textbf {\bibinfo
  {volume} {81}},\ \bibinfo {pages} {015803} (\bibinfo {year}
  {2010})}\BibitemShut {NoStop}%
\bibitem [{\citenamefont {Wang}\ \emph {et~al.}(2023)\citenamefont {Wang},
  \citenamefont {Ma}, \citenamefont {Chen}, \citenamefont {Ko}, \citenamefont
  {Sun},\ and\ \citenamefont {Zhang}}]{Wang:2023gta}%
  \BibitemOpen
  \bibfield  {author} {\bibinfo {author} {\bibfnamefont {R.}~\bibnamefont
  {Wang}}, \bibinfo {author} {\bibfnamefont {Y.-G.}\ \bibnamefont {Ma}},
  \bibinfo {author} {\bibfnamefont {L.-W.}\ \bibnamefont {Chen}}, \bibinfo
  {author} {\bibfnamefont {C.~M.}\ \bibnamefont {Ko}}, \bibinfo {author}
  {\bibfnamefont {K.-J.}\ \bibnamefont {Sun}},\ and\ \bibinfo {author}
  {\bibfnamefont {Z.}~\bibnamefont {Zhang}},\ }\bibfield  {title} {\bibinfo
  {title} {{Kinetic approach of light-nuclei production in intermediate-energy
  heavy-ion collisions}},\ }\href
  {https://doi.org/10.1103/PhysRevC.108.L031601} {\bibfield  {journal}
  {\bibinfo  {journal} {Phys. Rev. C}\ }\textbf {\bibinfo {volume} {108}},\
  \bibinfo {pages} {L031601} (\bibinfo {year} {2023})},\ \Eprint
  {https://arxiv.org/abs/2305.02988} {arXiv:2305.02988 [nucl-th]} \BibitemShut
  {NoStop}%
\bibitem [{\citenamefont {Kisiel}(2010)}]{Kisiel:2009eh}%
  \BibitemOpen
  \bibfield  {author} {\bibinfo {author} {\bibfnamefont {A.}~\bibnamefont
  {Kisiel}},\ }\bibfield  {title} {\bibinfo {title} {{Non-identical particle
  femtoscopy at s(NN)**(1/2) = 200-AGeV in hydrodynamics with statistical
  hadronization}},\ }\href {https://doi.org/10.1103/PhysRevC.81.064906}
  {\bibfield  {journal} {\bibinfo  {journal} {Phys. Rev. C}\ }\textbf {\bibinfo
  {volume} {81}},\ \bibinfo {pages} {064906} (\bibinfo {year}
  {2010})}\BibitemShut {NoStop}%
\bibitem [{\citenamefont {Vidana}\ \emph {et~al.}(2023)\citenamefont {Vidana},
  \citenamefont {Feijoo}, \citenamefont {Albaladejo}, \citenamefont {Nieves},\
  and\ \citenamefont {Oset}}]{Vidana:2023olz}%
  \BibitemOpen
  \bibfield  {author} {\bibinfo {author} {\bibfnamefont {I.}~\bibnamefont
  {Vidana}}, \bibinfo {author} {\bibfnamefont {A.}~\bibnamefont {Feijoo}},
  \bibinfo {author} {\bibfnamefont {M.}~\bibnamefont {Albaladejo}}, \bibinfo
  {author} {\bibfnamefont {J.}~\bibnamefont {Nieves}},\ and\ \bibinfo {author}
  {\bibfnamefont {E.}~\bibnamefont {Oset}},\ }\bibfield  {title} {\bibinfo
  {title} {{Femtoscopic correlation function for the Tcc(3875)+ state}},\
  }\href {https://doi.org/10.1016/j.physletb.2023.138201} {\bibfield  {journal}
  {\bibinfo  {journal} {Phys. Lett. B}\ }\textbf {\bibinfo {volume} {846}},\
  \bibinfo {pages} {138201} (\bibinfo {year} {2023})},\ \Eprint
  {https://arxiv.org/abs/2303.06079} {arXiv:2303.06079 [hep-ph]} \BibitemShut
  {NoStop}%
\bibitem [{\citenamefont {Liu}\ \emph {et~al.}(2025{\natexlab{b}})\citenamefont
  {Liu}, \citenamefont {Lu}, \citenamefont {Liu},\ and\ \citenamefont
  {Geng}}]{Liu:2024nac}%
  \BibitemOpen
  \bibfield  {author} {\bibinfo {author} {\bibfnamefont {Z.-W.}\ \bibnamefont
  {Liu}}, \bibinfo {author} {\bibfnamefont {J.-X.}\ \bibnamefont {Lu}},
  \bibinfo {author} {\bibfnamefont {M.-Z.}\ \bibnamefont {Liu}},\ and\ \bibinfo
  {author} {\bibfnamefont {L.-S.}\ \bibnamefont {Geng}},\ }\bibfield  {title}
  {\bibinfo {title} {{Femtoscopy can tell whether Zc(3900) and Zcs(3985) are
  resonances, virtual states, or bound states}},\ }\href
  {https://doi.org/10.1016/j.scib.2025.09.022} {\bibfield  {journal} {\bibinfo
  {journal} {Sci. Bull.}\ }\textbf {\bibinfo {volume} {70}},\ \bibinfo {pages}
  {3515} (\bibinfo {year} {2025}{\natexlab{b}})},\ \Eprint
  {https://arxiv.org/abs/2404.18607} {arXiv:2404.18607 [hep-ph]} \BibitemShut
  {NoStop}%
\bibitem [{\citenamefont {Acharya}\ \emph
  {et~al.}(2020{\natexlab{c}})\citenamefont {Acharya} \emph
  {et~al.}}]{ALICE:2019buq}%
  \BibitemOpen
  \bibfield  {author} {\bibinfo {author} {\bibfnamefont {S.}~\bibnamefont
  {Acharya}} \emph {et~al.} (\bibinfo {collaboration} {ALICE}),\ }\bibfield
  {title} {\bibinfo {title} {{Investigation of the
  p\textendash{}\ensuremath{\Sigma}0 interaction via femtoscopy in pp
  collisions}},\ }\href {https://doi.org/10.1016/j.physletb.2020.135419}
  {\bibfield  {journal} {\bibinfo  {journal} {Phys. Lett. B}\ }\textbf
  {\bibinfo {volume} {805}},\ \bibinfo {pages} {135419} (\bibinfo {year}
  {2020}{\natexlab{c}})},\ \Eprint {https://arxiv.org/abs/1910.14407}
  {1910.14407} \BibitemShut {NoStop}%
\bibitem [{\citenamefont {Acharya}\ \emph
  {et~al.}(2020{\natexlab{d}})\citenamefont {Acharya} \emph
  {et~al.}}]{ALICE:2020nkc}%
  \BibitemOpen
  \bibfield  {author} {\bibinfo {author} {\bibfnamefont {S.}~\bibnamefont
  {Acharya}} \emph {et~al.} (\bibinfo {collaboration} {ALICE}),\ }\bibfield
  {title} {\bibinfo {title} {{Multiplicity dependence of $\pi $, K, and p
  production in pp collisions at $\sqrt{s} = 13$ TeV}},\ }\href
  {https://doi.org/10.1140/epjc/s10052-020-8125-1} {\bibfield  {journal}
  {\bibinfo  {journal} {Eur. Phys. J. C}\ }\textbf {\bibinfo {volume} {80}},\
  \bibinfo {pages} {693} (\bibinfo {year} {2020}{\natexlab{d}})}\BibitemShut
  {NoStop}%
\bibitem [{\citenamefont {Abelev}\ \emph {et~al.}(2008)\citenamefont {Abelev}
  \emph {et~al.}}]{STAR:2008twt}%
  \BibitemOpen
  \bibfield  {author} {\bibinfo {author} {\bibfnamefont {B.~I.}\ \bibnamefont
  {Abelev}} \emph {et~al.} (\bibinfo {collaboration} {STAR}),\ }\bibfield
  {title} {\bibinfo {title} {{Hadronic resonance production in d+Au collisions
  at s(NN)**(1/2) = 200-GeV at RHIC}},\ }\href
  {https://doi.org/10.1103/PhysRevC.78.044906} {\bibfield  {journal} {\bibinfo
  {journal} {Phys. Rev. C}\ }\textbf {\bibinfo {volume} {78}},\ \bibinfo
  {pages} {044906} (\bibinfo {year} {2008})}\BibitemShut {NoStop}%
\bibitem [{\citenamefont {Adamczewski-Musch}\ \emph {et~al.}(2021)\citenamefont
  {Adamczewski-Musch} \emph {et~al.}}]{Adamczewski-Musch:2020edy}%
  \BibitemOpen
  \bibfield  {author} {\bibinfo {author} {\bibfnamefont {J.}~\bibnamefont
  {Adamczewski-Musch}} \emph {et~al.},\ }\bibfield  {title} {\bibinfo {title}
  {{Correlated pion-proton pair emission off hot and dense QCD matter}},\
  }\href {https://doi.org/10.1016/j.physletb.2021.136421} {\bibfield  {journal}
  {\bibinfo  {journal} {Phys. Lett. B}\ }\textbf {\bibinfo {volume} {819}},\
  \bibinfo {pages} {136421} (\bibinfo {year} {2021})}\BibitemShut {NoStop}%
\bibitem [{\citenamefont {Eskef}\ \emph {et~al.}(1998)\citenamefont {Eskef}
  \emph {et~al.}}]{FOPI:1998clk}%
  \BibitemOpen
  \bibfield  {author} {\bibinfo {author} {\bibfnamefont {M.}~\bibnamefont
  {Eskef}} \emph {et~al.} (\bibinfo {collaboration} {FOPI}),\ }\bibfield
  {title} {\bibinfo {title} {{Identification of baryon resonances in central
  heavy ion collisions at energies between 1-AGeV and 2-AGeV}},\ }\href
  {https://doi.org/10.1007/s100500050188} {\bibfield  {journal} {\bibinfo
  {journal} {Eur. Phys. J. A}\ }\textbf {\bibinfo {volume} {3}},\ \bibinfo
  {pages} {335} (\bibinfo {year} {1998})}\BibitemShut {NoStop}%
\bibitem [{\citenamefont {Hjort}\ \emph {et~al.}(1997)\citenamefont {Hjort}
  \emph {et~al.}}]{Hjort:1997wk}%
  \BibitemOpen
  \bibfield  {author} {\bibinfo {author} {\bibfnamefont {E.~L.}\ \bibnamefont
  {Hjort}} \emph {et~al.},\ }\bibfield  {title} {\bibinfo {title} {{Delta
  resonance production in Ni-58 + Cu collisions at E = 1.97-A-GeV}},\ }\href
  {https://doi.org/10.1103/PhysRevLett.79.4345} {\bibfield  {journal} {\bibinfo
   {journal} {Phys. Rev. Lett.}\ }\textbf {\bibinfo {volume} {79}},\ \bibinfo
  {pages} {4345} (\bibinfo {year} {1997})}\BibitemShut {NoStop}%
\bibitem [{\citenamefont {Reichert}\ \emph {et~al.}(2019)\citenamefont
  {Reichert}, \citenamefont {Hillmann}, \citenamefont {Limphirat},
  \citenamefont {Herold},\ and\ \citenamefont {Bleicher}}]{Reichert:2019lny}%
  \BibitemOpen
  \bibfield  {author} {\bibinfo {author} {\bibfnamefont {T.}~\bibnamefont
  {Reichert}}, \bibinfo {author} {\bibfnamefont {P.}~\bibnamefont {Hillmann}},
  \bibinfo {author} {\bibfnamefont {A.}~\bibnamefont {Limphirat}}, \bibinfo
  {author} {\bibfnamefont {C.}~\bibnamefont {Herold}},\ and\ \bibinfo {author}
  {\bibfnamefont {M.}~\bibnamefont {Bleicher}},\ }\bibfield  {title} {\bibinfo
  {title} {{Delta mass shift as a thermometer of kinetic decoupling in Au + Au
  reactions at 1.23 AGeV}},\ }\href {https://doi.org/10.1088/1361-6471/ab34fa}
  {\bibfield  {journal} {\bibinfo  {journal} {J. Phys. G}\ }\textbf {\bibinfo
  {volume} {46}},\ \bibinfo {pages} {105107} (\bibinfo {year}
  {2019})}\BibitemShut {NoStop}%
\bibitem [{\citenamefont {Reichert}\ \emph {et~al.}(2021)\citenamefont
  {Reichert}, \citenamefont {Hillmann},\ and\ \citenamefont
  {Bleicher}}]{Reichert:2020uxs}%
  \BibitemOpen
  \bibfield  {author} {\bibinfo {author} {\bibfnamefont {T.}~\bibnamefont
  {Reichert}}, \bibinfo {author} {\bibfnamefont {P.}~\bibnamefont {Hillmann}},\
  and\ \bibinfo {author} {\bibfnamefont {M.}~\bibnamefont {Bleicher}},\
  }\bibfield  {title} {\bibinfo {title} {{$\Delta$ resonances in Ca+Ca, Ni+Ni
  and Au+Au reactions from 1 AGeV to 2 AGeV: Consistency between yields, mass
  shifts and decoupling temperatures}},\ }\href
  {https://doi.org/10.1016/j.nuclphysa.2020.122058} {\bibfield  {journal}
  {\bibinfo  {journal} {Nucl. Phys. A}\ }\textbf {\bibinfo {volume} {1007}},\
  \bibinfo {pages} {122058} (\bibinfo {year} {2021})}\BibitemShut {NoStop}%
\bibitem [{\citenamefont {Weinhold}\ \emph {et~al.}(1996)\citenamefont
  {Weinhold}, \citenamefont {Friman},\ and\ \citenamefont
  {Noerenberg}}]{Weinhold:1996ts}%
  \BibitemOpen
  \bibfield  {author} {\bibinfo {author} {\bibfnamefont {W.}~\bibnamefont
  {Weinhold}}, \bibinfo {author} {\bibfnamefont {B.~L.}\ \bibnamefont
  {Friman}},\ and\ \bibinfo {author} {\bibfnamefont {W.}~\bibnamefont
  {Noerenberg}},\ }\bibfield  {title} {\bibinfo {title} {{Thermodynamics of an
  interacting pi N system}},\ }\href@noop {} {\bibfield  {journal} {\bibinfo
  {journal} {Acta Phys. Polon. B}\ }\textbf {\bibinfo {volume} {27}},\ \bibinfo
  {pages} {3249} (\bibinfo {year} {1996})}\BibitemShut {NoStop}%
\bibitem [{\citenamefont {Zhang}\ \emph {et~al.}(2026)\citenamefont {Zhang},
  \citenamefont {Shao}, \citenamefont {Zhang}, \citenamefont {Sun},\ and\
  \citenamefont {Ma}}]{Zhang:2026psh}%
  \BibitemOpen
  \bibfield  {author} {\bibinfo {author} {\bibfnamefont {L.}~\bibnamefont
  {Zhang}}, \bibinfo {author} {\bibfnamefont {T.}~\bibnamefont {Shao}},
  \bibinfo {author} {\bibfnamefont {S.}~\bibnamefont {Zhang}}, \bibinfo
  {author} {\bibfnamefont {K.-J.}\ \bibnamefont {Sun}},\ and\ \bibinfo {author}
  {\bibfnamefont {Y.-G.}\ \bibnamefont {Ma}},\ }\bibfield  {title} {\bibinfo
  {title} {{Resonance Femtoscopy Beyond the On-Shell Approximation}},\
  }\href@noop {} {\  (\bibinfo {year} {2026})},\ \Eprint
  {https://arxiv.org/abs/2606.08230} {arXiv:2606.08230 [nucl-th]} \BibitemShut
  {NoStop}%
\bibitem [{\citenamefont {Liu}\ \emph {et~al.}(2026)\citenamefont {Liu},
  \citenamefont {Lu}, \citenamefont {Ge},\ and\ \citenamefont
  {Geng}}]{Liu:2026vkc}%
  \BibitemOpen
  \bibfield  {author} {\bibinfo {author} {\bibfnamefont {Z.-W.}\ \bibnamefont
  {Liu}}, \bibinfo {author} {\bibfnamefont {J.-X.}\ \bibnamefont {Lu}},
  \bibinfo {author} {\bibfnamefont {D.-L.}\ \bibnamefont {Ge}},\ and\ \bibinfo
  {author} {\bibfnamefont {L.-S.}\ \bibnamefont {Geng}},\ }\bibfield  {title}
  {\bibinfo {title} {{Quantum interference effects enhanced in $\pi^+p$
  femtoscopic correlation functions}},\ }\href@noop {} {\  (\bibinfo {year}
  {2026})},\ \Eprint {https://arxiv.org/abs/2607.04351} {arXiv:2607.04351
  [hep-ph]} \BibitemShut {NoStop}%
\bibitem [{\citenamefont {Ropke}(2009)}]{Ropke:2008qk}%
  \BibitemOpen
  \bibfield  {author} {\bibinfo {author} {\bibfnamefont {G.}~\bibnamefont
  {Ropke}},\ }\bibfield  {title} {\bibinfo {title} {{Light nuclei quasiparticle
  energy shift in hot and dense nuclear matter}},\ }\href
  {https://doi.org/10.1103/PhysRevC.79.014002} {\bibfield  {journal} {\bibinfo
  {journal} {Phys. Rev. C}\ }\textbf {\bibinfo {volume} {79}},\ \bibinfo
  {pages} {014002} (\bibinfo {year} {2009})}\BibitemShut {NoStop}%
\bibitem [{\citenamefont {Sun}\ \emph {et~al.}(2019)\citenamefont {Sun},
  \citenamefont {Ko},\ and\ \citenamefont {D\"onigus}}]{Sun:2018mqq}%
  \BibitemOpen
  \bibfield  {author} {\bibinfo {author} {\bibfnamefont {K.-J.}\ \bibnamefont
  {Sun}}, \bibinfo {author} {\bibfnamefont {C.~M.}\ \bibnamefont {Ko}},\ and\
  \bibinfo {author} {\bibfnamefont {B.}~\bibnamefont {D\"onigus}},\ }\bibfield
  {title} {\bibinfo {title} {{Suppression of light nuclei production in
  collisions of small systems at the Large Hadron Collider}},\ }\href
  {https://doi.org/10.1016/j.physletb.2019.03.033} {\bibfield  {journal}
  {\bibinfo  {journal} {Phys. Lett. B}\ }\textbf {\bibinfo {volume} {792}},\
  \bibinfo {pages} {132} (\bibinfo {year} {2019})}\BibitemShut {NoStop}%
\end{thebibliography}

%

\begin{widetext}
\section{Supplemental material}
In the following, we give detailed derivations of $\pi^+-p$ and $\pi^+-d$ scattering cross sections, scattering wave functions, and correlation functions from a well established effective pion-nucleon interaction. We also analyze the effect of quantum interference between the  initial plane wave and the scattered $p-$wave.

\emph{Effective pion-nucleon interaction}--To describe the scattering between pion and nucleon, we adopt the following effective ($P$-wave)  $\pi N\Delta$ interaction Lagrangian  density~\cite{vanHees:2004vt},
\begin{equation}
\mathcal{L}
=
-\frac{f_\pi}{m_\pi}\,
\psi_N
\left[
(\vec S^{\dagger}\!\cdot\!\vec\nabla)
(\vec T^{\dagger}\!\cdot\!\vec\phi_\pi)
\right]
\Psi_\Delta
+ h.c.~. 
\end{equation}
In the above, $\vec\phi_\pi$, $
\psi_N$, and $\Psi_\Delta$ represent the pion, nucleon, and $\Delta$-resonance fields, respectively;  $f_\pi$ is the $\pi N\Delta$ coupling constant;  $m_\pi$ is the pion mass; and $\vec S$ and $\vec T$ are the spin 1/2 to spin 3/2 and isospin 1/2 to isospin 3/2 transition operators, respectively.

The amplitude for the process $\pi+ N\rightarrow\Delta\rightarrow \pi+N$ is then given by
\begin{equation}
\begin{aligned}
i\mathcal{M}=-i \frac{f_\pi^2}{m_\pi^2}\,\sqrt{2E_N(k)2E_N(k')}\,
\frac{F(k)F(k')}{\sqrt{s}-m_\Delta+i\Gamma(k)/2}
\left(\frac{2}{3}\boldsymbol{k}\cdot \boldsymbol{k}'
-\frac{i}{3}\bm{\sigma}\cdot(\boldsymbol{k}'\times \boldsymbol{k})\right),
\end{aligned}
\end{equation}
where $\boldsymbol{k}$ and $\boldsymbol{k}'$ are the initial and final pion three-momenta in the center-of-mass (CM) frame, respectively, with $k=|\boldsymbol{k}|$ and $k'=|\boldsymbol{k}'|$, and $F(k)=\Lambda^2/(\Lambda^2+k^2)$ is the form factor at the $\pi N\Delta$ vertex~\cite{vanHees:2004vt}.  Averaging the initial spin and summing over the final spin gives
\begin{equation}
\begin{aligned}
\overline{|\mathcal{M}|^2}
= \left(\frac{f_\pi}{m_\pi}\right)^4
4E_N(k)^2
\frac{F(k)^4}{(\sqrt{s}-m_\Delta)^2+\Gamma(k)^2/4}
\frac{k^4}{9}\left(1+3\cos^2\theta\right),
\end{aligned}
\end{equation}
where $\theta$ denotes the azimuthal angle between ${\bf k}$ and ${\bf k^\prime}$. The differential cross-section in the CM frame is then given by 
\begin{equation}
 \frac{d\sigma}{d\Omega}  = \frac{1}{2E_N (k)2E_{\pi}(k) |v_N - v_{\pi}|} \frac{k}{(2\pi)^2 4\sqrt{s}} \overline{|\mathcal{M}|^2}.
\end{equation}
 Integrating over the solid angle yields the total $\pi^+-p$ cross section  as
\begin{equation}
\sigma_{\pi^+-p} = \frac{8\pi }{k^2} \frac{\Gamma(k)^2/4}{\bigl(\sqrt{s} - m_\Delta\bigr)^2 + \Gamma(k)^2/4}.
\end{equation}
The decay width in above equations can be calculated from the same interaction Lagrangian density, and it is given by~\cite{vanHees:2004vt}
\begin{equation}
\Gamma(k) =  \frac{f_\pi^2}{6\pi m_\pi^2} \frac{E_N(k)k^3}{E_\pi(k)+E_N(k)}F^2(k).
\end{equation}
Using $m_{\pi}=0.139$ GeV, $m_{N}=0.939$ GeV, $m_{\Delta}=1.23$ GeV,  $f=3.3$ and $\Lambda=0.33$ GeV, the above cross section reproduces well the measured cross section~\cite{ParticleDataGroup:2020ssz} as shown in window (a) of Fig.~\ref{fig:cross_section}.

For $\pi^+-d$ scattering, its total cross section under the impulse approximation is given by
\begin{equation}
\sigma_{\pi^+-d}\approx\sigma_{\pi^+-p}+\sigma_{\pi^+-n}\approx\frac{4}{3}\sigma_{\pi^+-p}.
\label{eq:cs-pid}
\end{equation}
As displayed in Fig.~\ref{fig:cross_section} (b), the cross section  obtained from the impulse approximation describes reasonably the experimental data~\cite{ParticleDataGroup:2020ssz}. 

\begin{figure}[!t]
\centering
\includegraphics[width=0.8\textwidth]{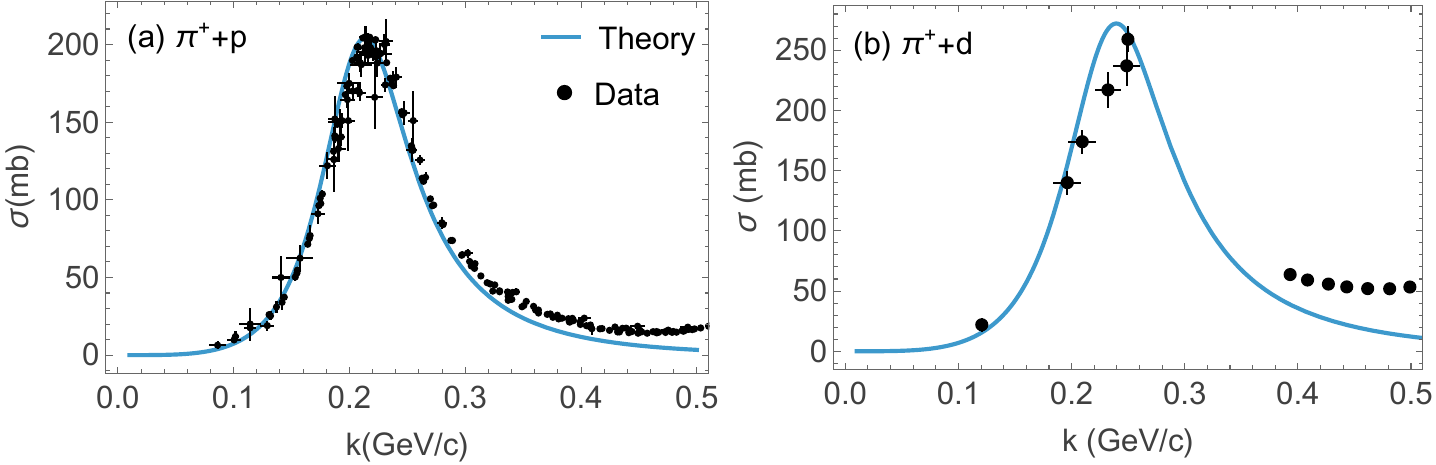}
\caption{Theoretical cross section in comparison to experimental data~\cite{ParticleDataGroup:2020ssz}.} 
\label{fig:cross_section}
\end{figure}

\emph{Scattering wave function of $\pi^+-p$.}--
The full $\pi^+-p$ scattering wave function in coordinate space  satisfies the Lippmann-Schwinger equation~\cite{Vidana:2023olz,Liu:2024nac}:
\begin{equation}
\Psi_{\boldsymbol{k}}(\boldsymbol{r}) = e^{i \boldsymbol{k}\cdot \boldsymbol{r}} + \int \frac{d^{3}{\boldsymbol{k}'}}{(2\pi)^3} G(k,k')T(\boldsymbol{k}',\boldsymbol{k})e^{i \boldsymbol{k}' \cdot \boldsymbol{r}},
\end{equation}
where the $\pi^+-p$ propagator is given by 
\begin{equation}
G(k,k') =\frac{m_\pi m_N}{E_\pi(k') E_N(k')} \frac{1}{E_{N}(k) + E_{\pi}(k) - E_N(k') - E_{\pi}(k') + i\epsilon},
\end{equation}
and the $\pi^+-p$ scattering $T$-matrix is
\begin{equation}
\begin{aligned}
T(\boldsymbol{k}',\boldsymbol{k})={}&\frac{f_\pi^2}{m_\pi^2}
\frac{1}{\sqrt{2E_\pi(k)\,2E_\pi(k')}}
\frac{F(k)F(k')}
{E_N(k)+E_\pi(k)-m_\Delta+i\Gamma (k)/2}\\
&\times\left(
\frac{2}{3}\boldsymbol{k}\cdot \boldsymbol{k}'
-\frac{i}{3}\boldsymbol{\sigma}\cdot
(\boldsymbol{k}'\times \boldsymbol{k})
\right).
\end{aligned}
\end{equation}

Solving the Lippman-Schwinger equation gives the following $\pi^+ - p$ scattering wave function in coordinate space: 
\begin{equation}
\Psi_{\boldsymbol{k}}(\boldsymbol{r}) = e^{i \boldsymbol{k} \cdot \boldsymbol{r}} + \left[ \frac{2}{3} (\hat{\boldsymbol{k}} \cdot \hat{\boldsymbol{r}}) - \frac{i}{3} \boldsymbol{\sigma} \cdot (\hat{\boldsymbol{r}} \times \hat{\boldsymbol{k}}) \right]  U(k,r),
\end{equation}
where 
\begin{equation}\label{eq:U}
U(k,r) = \frac{ik}{2\pi^2} B(k) \int_0^\infty dk' \frac{m_\pi m_N}{E_\pi(k') E_N(k')}\frac{(k')^3 F(k') j_1(k'r)}{\sqrt{2E_\pi(k')}(E_N(k) + E_\pi(k) - E_N(k') - E_\pi(k') + i\epsilon)},
\end{equation}
with $j_1(k'r)$ being the spherical Bessel function, and
\begin{equation}
B(k) = \frac{f_{\pi}^2}{m_\pi^2} \frac{F(k)}{\sqrt{2E_{\pi}(k)} \left( E_N(k) + E_\pi(k) - m_\Delta + i\Gamma(k)/2 \right)}.
\end{equation}

For  initial nucleon in spin states $\chi_\pm$, its spin-dependent scattering wave functions are:
\begin{equation}
\Psi_{\boldsymbol{k}}(\boldsymbol{r})\chi_+ = e^{i\boldsymbol{k}\cdot\boldsymbol{r}}\chi_+ + U(k,r)\left(\frac{2}{3}\cos\theta\chi_+ - \frac{1}{3}\sin\theta e^{i\phi}\chi_-\right),
\end{equation}
\begin{equation}
\Psi_{\boldsymbol{k}}(\boldsymbol{r}) \chi_{-} = e^{i\boldsymbol{k}\cdot\boldsymbol{r}} \chi_{-} + U(k,r) \left( \frac{2}{3} \cos \theta \chi_{-} + \frac{1}{3} \sin \theta e^{-i\phi} \chi_{+} \right).
\end{equation}
With the four spin matrix elements given by
\begin{equation}
\begin{aligned}
\langle \chi_+ | \Psi | \chi_+ \rangle &= e^{i\boldsymbol{k}\cdot\boldsymbol{r}} + \frac{2}{3} U \cos \theta, &\qquad \langle \chi_- | \Psi | \chi_+ \rangle &= - \frac{1}{3} U \sin \theta e^{i\phi}, \\
\langle \chi_- | \Psi | \chi_- \rangle &= e^{i\boldsymbol{k}\cdot\boldsymbol{r}} + \frac{2}{3} U \cos \theta, &\qquad \langle \chi_+ | \Psi | \chi_- \rangle &= \frac{1}{3} U \sin \theta e^{-i\phi},
\end{aligned}
\end{equation}
summing over final spin states and averaging over initial spin states  leads to 
\begin{equation}
\begin{aligned}
\overline{|\Psi|^2}
=1+|U(k,r)|^2
\left(\frac{1}{3}\cos^2\theta+\frac{1}{9}\right)
+\frac{4}{3}\cos\theta\,
\operatorname{Re}\!\left[
U(k,r)e^{-i\boldsymbol{k}\cdot\boldsymbol{r}}
\right].
\end{aligned}
\end{equation}

\emph{Coulomb correction.}--Within the modified Lippmann-Schwinger formalism based on the two-potential framework, the 
effect of Coulomb potential can be included by replacing the free plane wave $e^{i\boldsymbol{k}\cdot\boldsymbol{r}}$  with the exact  pure Coulomb wave function,
\begin{equation}
e^{i\boldsymbol{k}\cdot\boldsymbol{r}} \rightarrow \Psi_{0}^{C}(r,k,\cos\theta)=e^{-\pi\eta/2} \, \Gamma(1+i\eta) \, e^{i\boldsymbol{k}\cdot\boldsymbol{r}} \, {}_1F_1\left(-i\eta; 1; i(kr -\boldsymbol{k}\cdot\boldsymbol{r})\right),
\end{equation}
where  the dimensionless Sommerfeld (Gamow) parameter $\eta$ is defined as:
\begin{equation}
\eta(k) = \frac{\alpha \, E_\pi(k) \, E_N(k)}{k \left(E_\pi(k) + E_N(k)\right)},
\end{equation}
with $\alpha \approx 1/137$ being the fine-structure constant. This leads to the modification of Eq. (\ref{eq:U}) to 
\begin{equation}
U(k,r) \rightarrow \frac{ik}{2\pi^2} B(k) e^{i\sigma_1} \int_0^\infty dk' \frac{m_\pi m_N}{E_\pi(k') E_N(k')}\frac{k'^3 \, F(k') \, \left[\frac{F_1(\eta(k'), k'r)}{k'r}\right]}{\sqrt{2E_\pi(k')}\left(E_N(k) + E_\pi(k) - E_N(k') - E_\pi(k') + i\epsilon\right)}.
\end{equation}
where $\sigma_\ell=\arg\Gamma(\ell+1+i\eta)$ is the Coulomb phase shift, with $\Gamma$ denoting the Euler gamma function, while $F_\ell(\eta,\rho)=\frac{2^\ell e^{-\pi\eta/2}|\Gamma(\ell+1+i\eta)|}{(2\ell+1)!}
\allowbreak \rho^{\ell+1}e^{i\rho} \allowbreak M(\ell+1+i\eta,2\ell+2,-2i\rho)$ is the regular Coulomb wave function, where $M$ denotes the confluent hypergeometric function of the first kind.

\emph{Scattering wave function $\pi^+-d$.}--
For $\pi^+-d$ scattering, the problem is in principle a three-body one. However, within the impulse approximation, it can be treated as the scattering of the pion off a nucleon inside the deuteron. Neglecting the small binding energy  and taking the proton momentum inside the deuteron  to be half of the deuteron momentum, the $U(k,r)$ in Eq.~(\ref{eq:U}) is replaced by $\sqrt{4/3}U(\tilde{k},\tilde{r})$, where the  factor $\sqrt{4/3}$ accounts for the cross section relation $\sigma_{\pi^+-d}\approx\sigma_{\pi^+-p}+\sigma_{\pi^+-n}\approx\frac{4}{3}\sigma_{\pi^+-p}$  in the impulse approximation. The  $\tilde{k}$ and $\tilde{r}$  in the argument of $U(\tilde k, \tilde r)$ denote, respectively, the relative momentum and  coordinate between the pion and the proton inside the deuteron.  

The relative momentum $\tilde k$ can be determined from the squared  invariant mass   $s_{\pi p}(k_{\pi d})$   of  the pion and the proton inside the deuteron  given by
\begin{equation}
 s_{\pi p}(k_{\pi d})
 =
 m_{\pi}^{2}+m_{p}^{2}
 +
 \sqrt{m_{\pi}^{2}+k_{\pi d}^{2}}
 \sqrt{(2m_{p})^{2}+k_{\pi d}^{2}}
 +k_{\pi d}^{2}, 
 \label{eq:s-piN-pid-rel-2mN}
\end{equation}
where $k_{\pi d}$ denotes the momentum of  the  pion in the center-of-mass frame of  the  pion-deuteron pair. The relative momentum in the $\pi p$ center-of-mass frame is  then give by  
\begin{equation}
\tilde{k}
 =
 \frac{
 \sqrt{
 \lambda\!\left(
 s_{\pi p}(k_{\pi d}),
 m_{\pi}^{2},
 m_{p}^{2}
 \right)}
 }{
 2\sqrt{s_{\pi p}(k_{\pi d})}
 },
 \label{eq:p-piN-pid-rel}
\end{equation}
with
$\lambda(a,b,c)
 =
 a^{2}+b^{2}+c^{2}-2ab-2ac-2bc.
$

For the relative coordinate $\tilde{r}$ between the pion and the proton in the deuteron, we take it  to be the relative distance between pion and deuteron, i.e., neglecting the smearing effect by deuteron wavefunction. 

\begin{figure}[!t]
\centering
\includegraphics[width=0.9\textwidth]{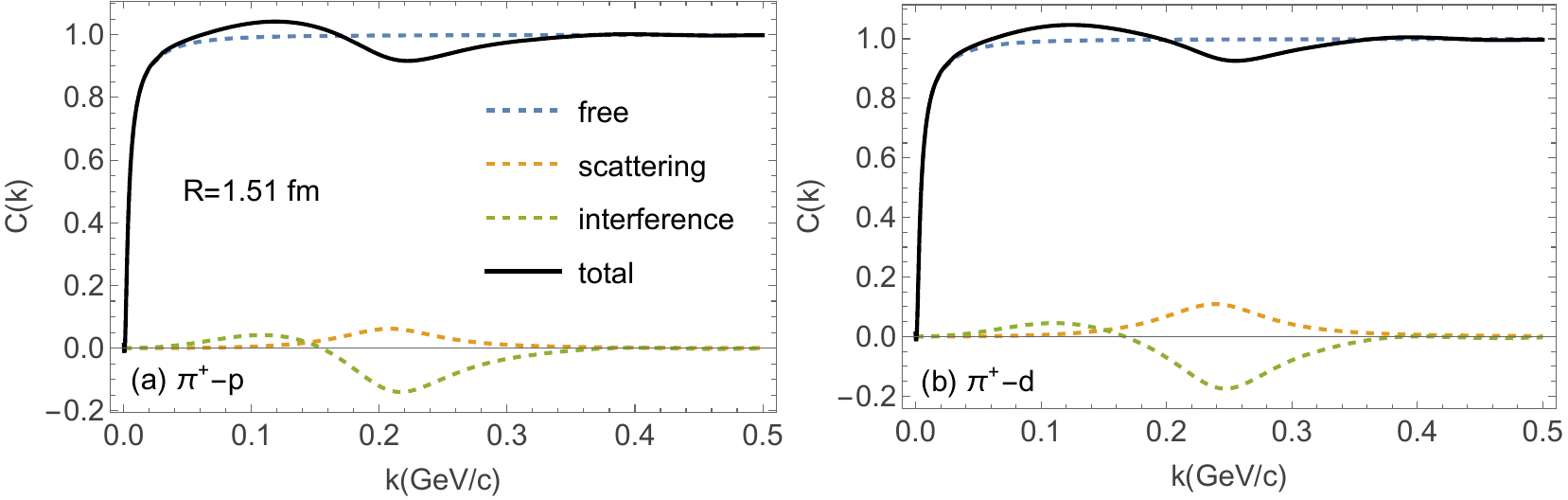} 
\caption{Correlation functions
$C$ of $\pi^+-p$ (a) and $\pi^+-d$ (b) for a Gaussian source with $R=1.51~\mathrm{fm}$ in the case of pure $p-$wave scattering with Coulomb interaction.} 
\label{fig:correlation_function}
\end{figure}

\emph{Correlation function with an isotropic Gaussian source.}{\bf ---} For  an emission source of Gaussian distribution in coordinate space, $S(r) = (4\pi R^2)^{-3/2} \exp(-r^2/4R^2)$, the total correlation function  can be decomposed into
\begin{equation} 
C(k)  = C_{\text{free}}(k) + C_{\text{scat}}(k) + C_{\text{int}}(k),\label{eq:corr1}
\end{equation}
with
\begin{equation}
\begin{aligned} 
C_{\text{free}}(k) &= 2\pi
 \int_{0}^{\infty}dr\,r^{2}S(r)
 \int_{-1}^{1}d(\cos\theta)\,
 \left|
 \Psi_{0}^{C}(r,k,\cos\theta)
 \right|^{2}, \\[4pt]
C_{\text{scat}}(k) &= \frac{8\pi}{9} \int_0^\infty r^2 S(r) |U(k,r)|^2 \, dr, \\[4pt]
C_{\text{int}}(k) &= \int_0^\infty r^2 S(r) \, \text{Re} \left[ -\frac{16\pi i}{3} j_1(kr) U(k,r) \right] dr \\
&= \frac{16\pi }{3}\int_0^\infty r^2 S(r) \, \text{Im} \left[  j_1(kr) U(k,r) \right] dr,
\end{aligned}\label{eq:corr}
\end{equation}
In the above, $C_{\text{free}}$ represents the contribution from pure plane wave with Coulomb repulsion, $C_{\text{scat}}$ denotes contribution purely from scattered $p-$wave, and $C_{\text{int}}$ denotes the contribution from their quantum interference. As shown in Fig.~\ref{fig:correlation_function}~(a), $C_{\text{scat}}$ peaks near the $\Delta$-resonance mass, whereas the peak–dip structure of $C_{\text{int}}$ shifts the peak position to lower momentum. The Coulomb repulsion primarily suppresses $C_{\text{free}}$ at very low $k<0.05$ GeV/$c$. 

\begin{figure}[!h]
\centering
\includegraphics[width=0.9\textwidth]{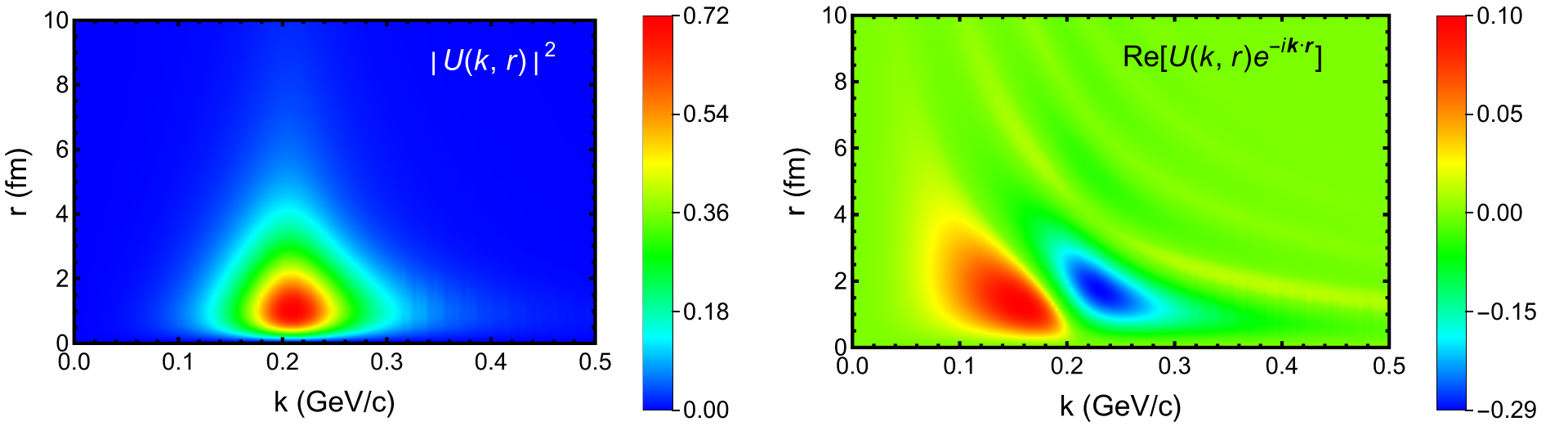} 
\caption{Contour plots of $|U(k,r)|^2$ (a) and $\text{Im}[j_1(kr)U(k,r)]$ (b) in the $r-k$ plane.} 
\label{fig:interference}
\end{figure}

\emph{Effect of interference.}{\bf ---}To  illustrate the effect of the interference term in Eq.~(\ref{eq:corr}), we show in Fig.~\ref{fig:interference} contour plots for   $|U(k,r)|^2$ (panel (a)) and $\text{Im}[j_1(kr)U(k,r)]$ (panel (b))  in the $r-k$ plane for $\pi^+-p$ scattering. It is seen that $|U(k,r)|^2$ peaks at about $k\approx 0.22$ GeV/$c$ for any $r$, while $\text{Im}[j_1(kr)U(k,r)]$ peaks at a much smaller $k$ and also shows a large dip at $k=0.2-0.3$ GeV/$c$. The  $\text{Im}[j_1(kr)U(k,r)]$ exhibits an oscillator behavior in the $r-k$ plane expected from the quantum interference. However, this interference effect is so strong that the peak in $C_\text{scat}$ is insufficient to compensate for the dip in $C_\text{int}$, resulting in a total correlation function that remains to have a low-$k$ peak and a high-$k$ dip. Only after including the contribution from direct $\Delta$ resonance decay to the correlation function, this dip can fully disappear~\cite{Liu:2026vkc}.

\end{widetext} 

\end{document}